\documentclass[11pt]{article}

\usepackage[dvips]{graphicx}
\usepackage{latexsym,amsmath,amsfonts,amscd}
\usepackage{epsfig}
\usepackage{changebar}
\usepackage{multirow}
\usepackage{verbatim}
\usepackage{graphicx, graphics}
\usepackage{amsmath}
\usepackage{amssymb}
\usepackage{color}

\usepackage{algorithmic}  
\usepackage[ruled,vlined]{algorithm2e}

\newcommand{\bk}{\mathbf{k}}
\newcommand{\bx}{\mathbf{x}}

\newcommand{\eps}{\varepsilon}
\newcommand{\bE}{\mathbf{E}}
\newcommand{\ti}{\mathrm{t}}
\newcommand{\nV}{\Psi}
\newcommand{\kbt}{k_B T_L}

\newcommand{\hw}{\hbar \omega_{p}}
\newcommand{\mass}{m^{*}}

\newfont{\iams}{msbm9}

\newcommand{\ot}{\frac{1}{2}}
\newcommand{\dm}{\displaystyle}

\newcommand{\cd}[1]{c_{#1}}
\newcommand{\alp}{\alpha_{p}}
\newcommand{\domk}{\Omega_{\bk}}
\newcommand{\intd}{\int_{\domk}}

\newcommand{\enk}{\eps(\bk)}

\newcommand{\bka}{\bk'}

\newcommand{\enka}{\eps(\bka)}

\newcommand{\nq}{{\sf n}_{q}}

\newcommand{\kB}{k_{B}}
\newcommand{\TL}{T_{L}}
\newcommand{\kT}{\kB \TL}

\newcommand{\ib}{\bar{i}} 
\newcommand{\jb}{\bar{j}} 
\newcommand{\kb}{\bar{k}} 
\newcommand{\mb}{\bar{m}} 
\newcommand{\nb}{\bar{n}} 
\begin{document}

\baselineskip=2pc

\begin{center}
{\bf Discontinuous Galerkin Deterministic Solvers for a
Boltzmann-Poisson Model of Hot Electron Transport by Averaged
Empirical Pseudopotential Band Structures
\footnote{Support from the Institute of Computational
Engineering and Sciences and the University of Texas Austin is
gratefully acknowledged.}}
\end{center}

\vspace{.10in}

\centerline{ Jos\'e Morales-Escalante \footnote{E-mail: jmorales@ices.utexas.edu}, 
Irene M. Gamba\footnote{E-mail: gamba@math.utexas.edu. 
JME and IMG  are partially supported by NSF DMS 1109625, NSF  CHE-0934450, NSF-RNMS  DMS-1107465  and the ICES Moncrief  Grand Challenge Award. }}

\smallskip

\centerline{ICES and Department of Mathematics, University of Texas,
Austin, TX 78712}

\bigskip

\centerline{
Yingda Cheng\footnote{E-mail: ycheng@math.msu.edu. 
Research supported by NSF grants DMS-1318186 and DMS-1453661.}}
 
\smallskip
 
\centerline{Department of Mathematics, Michigan State University,
East Lansing, MI 48824}

\bigskip

\centerline{Armando Majorana\footnote{E-mail:majorana@dii.unict.it. Research supported by J.T. Oden Visitors Program, ICES, The University of Texas at Austin.}}

\smallskip

\centerline{Department of Industrial Engineering,
University of Catania, Catania, Italy}

\bigskip

\centerline{Chi-Wang Shu\footnote{E-mail: shu@dam.brown.edu.  Research
supported by DOE grant DE-FG02-08ER25863 and NSF grant DMS-1418750.}}

\smallskip

\centerline{Division of Applied Mathematics, Brown University,
Providence, RI 02912}

\centerline{\em and}

\smallskip

\centerline{ 
James Chelikowsky \footnote{E-mail: jrc@ices.utexas.edu. JRC acknowledges support provided by the Scientific Discovery
through Advanced Computing (SciDAC) program funded by U.S.~DOE, Office
of Science,
Advanced Scientific Computing Research and Basic Energy Sciences under
Award No. DE-SC0008877. }}

\smallskip

\centerline{Department of Physics and ICES, University of Texas,
Austin, TX 78712}

\bigskip

\centerline{February 2017}

\newpage

\vspace{.4in}

\centerline{\bf Abstract}

\vspace{.16in}

The purpose of this work is to incorporate numerically, in a discontinuous Galerkin (DG) solver of a 
Boltzmann-Poisson model for hot electron transport, an electronic conduction band whose values are 
obtained by the spherical averaging of the full band structure given by a 
local empirical 
pseudopotential method (EPM) around a local minimum of the conduction 
band for silicon, as a 
midpoint between a radial band model and an anisotropic full band, in order to provide a more 
accurate physical description of the electron group velocity and conduction energy band structure in 
a semiconductor. 
This gives a better quantitative description of the transport and collision phenomena
that fundamentally define the behaviour of the Boltzmann - Poisson model for electron transport used in this work. 
The numerical values of the derivatives of this conduction energy band, needed for the 
description of the electron group velocity, are obtained by means of a 
cubic spline interpolation. 
The EPM-Boltzmann-Poisson transport with this spherically averaged EPM 
calculated energy surface is 
numerically simulated and compared to the output of traditional analytic 
band models such as the 
parabolic and Kane bands, numerically implemented too, for the case of 
1D $n^+-n-n^+$ silicon 
diodes with $400nm$ and $50nm$ channels. Quantitative differences are
observed in the kinetic moments related to the conduction energy band 
used, such as mean velocity, average energy, and electric current (momentum).

\vfill


\vfill

\newpage

\section{Introduction}
The Boltzmann-Poisson (BP) system is a semi-classical model for electric charge transport in semiconductors. 
The BP system can be used to describe the hot electron transport in modern 
semiconductor devices at nano-scales. As stated in \cite{ref:MarkowichRS}, this model describes the long range interactions over charge carriers 
and the statistical evolution of its states that includes an account of the
quantum scattering events. The BP system treats charge carriers partly as classical particles by describing 
them by means of a time-dependent probability density function $f(\ti,\bx,\bk)$ over the phase space 
$(\bx,\bk)$, and using a Boltzmann equation to model the time evolution of 
the associated 
probability density function in the phase space. 
The quantum nature of the carriers is considered in 
several terms of the Boltzmann equation. 
The quantum crystal wave-vector $\bk$ is used as the momentum phase space variable in the model.
The model for the local velocity of the charge carriers is their respective group velocity 
$\mathbf{v}(\bk) = \frac{1}{\hbar} \nabla_{\bk}\enk$, related to the electronic energy band function 
$\enk$ of the considered semiconductor material. 
As usual, $\hbar$ is the Planck constant divided by $2\pi$.
The collision integral operator models 
the quantum scattering mechanisms acting over the charge carriers. 
The flow of charge carriers is induced by the force over the electron charge $-q$, which is assumed 
to be given by a mean electric field,  $\mathbf{F}(\ti,\bx) = -q \, \bE(\ti,\bx)$. This 
effective electric field, modeled by the Poisson Equation, 
takes into account long range interactions made of both internal carrier 
self-consistent and external contributions, such as an applied potential (bias). 
Hence, time-dependent solutions of the the BP system contain all the information on the transient of the 
carrier distribution and the time evolution of the total electric field. A phenomenological 
derivation of the BP model can be found in \cite{ref:MarkowichRS}.

The semi-classical Boltzmann description of electron transport in semiconductors is, for a truly 3-D 
device, an equation in six dimensions plus time when the device is not in steady state. For a 1-D 
device model, under azimuthal symmetry assumptions, the dimensionality of the problem can be reduced to 3 
dimensions plus time. The  heavy computational cost is the main reason why the BP system had been 
traditionally solved numerically by means of Direct Simulation Monte Carlo (DSMC) methods 
\cite{jaco89}. However, after the pioneer work \cite{Fatemi_1993_JCP_Boltz}, in recent years, 
deterministic solvers to the BP system were proposed in 
\cite{MP, carr02, cgms03, carr03, carr021, cgms06, gm07, galler-re}. 
These methods provide accurate results which, in general, agree well with those obtained from Monte 
Carlo (DSMC) simulations, often at a fractional computational time.  Moreover, these type of solvers can resolve 
transient details for the electron probability density function $f$, which are 
difficult to compute with DSMC simulators. 
The initial methods proposed in \cite{cgms03, carr03, carr021, cgms06} using weighted essentially non-oscillatory (WENO) finite 
difference schemes to solve the Boltzmann-Poisson system, had the advantage that the scheme is relatively simple to code and very stable even on coarse meshes for solutions containing sharp 
gradient regions. However, a disadvantage of the WENO methods is that it requires smooth meshes to achieve high order accuracy, hence it is not very flexible for adaptive meshes.

Motivated by the easy {\it hp-}adaptivity ({\it h} refers to the size of the element, and {\it p} refers to the polynomial degree of the space generated by the basis functions) and the simple communication pattern of 
the discontinuous Galerkin (DG) methods for macroscopic (fluid level) models \cite{Chen_95_JCP_Q, 
Chen_95_VLSI, LS1,LS2}, it was proposed in \cite{chgms-sispad07, Cheng_08_JCE_BP} to implement a DG 
solver to the full Boltzmann equation, that is capable of capturing transients of the probability 
density function. 
In the previous work \cite{chgms-sispad07,Cheng_08_JCE_BP}, the first DG solver for the BP system
was proposed, and some numerical calculations were shown for one and 
two-dimensional devices.  In \cite{CGMS-CMAME2008}, the DG-LDG scheme for the Boltzmann-Poisson 
system was carefully formulated, and extensive numerical studies were performed to validate the 
calculations. Such scheme 
models electron transport along the conduction band for 1D diodes and 2D double gate MOSFET devices 
with the energy band $\enk = \varepsilon( | \bk | )$ given by the Kane band model (valid close to a 
local minimum) in which the relation between the energy $\varepsilon$ and the wavevector norm $| \bk 
|$ is given by the analytic formula, referred as the Kane band model: 
\begin{equation}
\label{kaneband}
\varepsilon \left( 1 + \alpha \varepsilon \right) = \frac{ \hbar^2 |\bk|^2 }{2m^*} 
\end{equation}
where $m^*$ is the effective mass for the considered material, Silicon for the case of this paper, 
and $\alpha$ is a non-parabolicity constant. This band model can be understood as a first order 
variation from the parabolic band model, given by the particular case
$\alpha = 0 $. 

In all of the aforementioned deterministic solvers previous to \cite{CGMS-IWCE13}, the energy-band  function $\enk$  is given analytically, either by  the parabolic band approximation or by the Kane non-parabolic band model. 
The analytical band makes use of the explicit dependence of 
the carrier energy on the quasimomentum, which significantly simplifies all expressions as well as 
implementation of these techniques in the collision operator. However, some physical details of the 
band structure are partly or totally ignored when using an analytic approximation, which hinders its 
 application to transport of hot carriers in high-field phenomena (the so called hot electron 
transport) where the high anisotropy of the real band structure far from the conduction band minimum 
becomes important. Full band models, on the other hand, are able to provide an accurate physical 
description of the energy-band function, portraying this anisotropic band structure far from a 
conduction band minimum. 
One of the most commonly used methods to compute 
full bands is the 
empirical pseudopotential method (EPM). Such method gives a full 
band structure truncating the Fourier 
series in the $k$-space \cite{paperCHLKW} for a crystal lattice potential model given as the sum of 
potentials due to individual atoms and associated electrons, with few parameters fitting empirical 
data such as optical gaps, absorption rates, etc, to finally compute the energy eigenvalues of the 
Schr\"{o}dinger equation in 
Fourier space. A more detailed discussion of this method can be found in 
\cite{paperCHLKW, COHEN-CHLKW}. 
While full band models, as the ones given by EPM, have been widely 
used in DSMC simulators \cite{jaco89}, their inclusion in deterministic solvers for the transport Boltzmann Equation is more recent; on \cite{Smirnov_06_JAP_Fullband}, \cite{FBvalence}, full band models have also been combined with spherical harmonic expansion methods used to solve the Boltzmann equation numerically. However, high order accuracy is not always achieved by spherical harmonic expansion methods 
when energies vary strongly
and only a few terms of the expansion are usually employed \cite{Majorana_2002}. In contrast, the simulations for the BP system  developed in our line of work, as in 
\cite{chgms-sispad07}, \cite{CGMS-CMAME2008}, do not involve any asymptotics
and so are very accurate for hot electron transport regimes.
A DG method for full conduction bands BP models was proposed 
in \cite{CGMS-IWCE13},
generalizing the solver that uses the Kane non-parabolic band and adapting it to treat the full energy band case. A preliminary benchmark of numerical results shows that the the Dirac delta functional in the energy variable can be applied in this case to reduce one dimension of the collision integral, and so an accurate high-order simulation with comparable computational cost to the analytic band 
cases is possible.

The work presented in this paper is focused on simulations for hot electron transport along a single 
conduction band for Si computed by radially averaging an EPM full band structure. The band obtained by this procedure represents a midpoint 
between a radial band model and a full band anisotropic model, having then both the desired advantages 
of a band model with a dependence on $r$ and, at the same time having information of the 
anisotropic variation of the conduction band in the
$\bk$-space, by means of the numerical average 
performed over the angular domain of the conduction band for a given 
$\bk$-sphere. The advantage of DG scheme in this framework is shown in the accurate calculation of the Dirac delta functions in the collisional integrals based on the weak formulation. The EPM-Boltzmann-Poisson transport with this spherically averaged EPM 
calculated energy surface is 
numerically simulated and compared to the output of traditional analytic 
band models such as the 
parabolic and Kane bands. Quantitative differences are
observed in the  moments, demonstrating the significance of incorporating the physical band models in the kinetic simulations.


%


The rest of the paper is 
organized as follows: in Sections 2 and 3, we present the BP model, and the transformed equations under spherical coordinates in $\bk$. Section 4 contains the details of the computations of the spherical average of a local EPM conduction band. The DG formulation is presented in Section 5.  The device configuration and numerical results are discussed in Sections 6 and 7. We conclude the paper in Section 8. Some technical details of the schemes are given in the Appendix.
\section{The Boltzmann-Poisson problem}

We consider the probability density function (\emph{pdf}) for electrons along a single conduction band, denoting it by $f(t,\bx,\bk)$
We denote by $\Omega_{\bx}$ the physical domain in the $\bx$-space, and similarly $\domk$ as the domain in the $\bk$-space.
Following \cite{ref:MarkowichRS}, we recall the \emph{classical (strong) formulation of the initial value problem for the BP system with boundary 
conditions, for a pdf of electrons on a single conduction band:} 

Find $f: \mathbb{R}^{+} \times \Omega_{\bx} \times \domk \rightarrow  \mathbb{R}$, $\quad 
f(\ti,\bx,\bk) 
\geq 0$ and $V(t, \bx): \mathbb{R}^{+} \times \Omega_{\bx} \rightarrow  \mathbb{R}$, 
such that the Boltzmann equation
$$
\frac{\partial f}{\partial \ti} + \frac{1}{\hbar} \nabla_{\bk} \,
\varepsilon (\bk) \cdot \nabla_{\bx} f -
 \frac{q}{\hbar} \bE (\ti,\bx) \cdot \nabla_{\bk} f = Q(f) \, ,
$$
with the linear collision operator $Q(f)$ describing the 
scattering over the electrons,
where several quantum mechanisms can be taken into account.
In the low density approximation, the collisional 
integral operator becomes linear in $f$, having the form:
\begin{equation}
Q(f) = \intd \left[ S(\bk', \bk) f(\ti, \bx, \bk') - S(\bk, \bk') f(\ti, \bx, \bk) \right] d \bk'
\label{ope_coll}
\end{equation}
where $S(\bk,\bk')$ is the scattering kernel,
representing non-local interactions of electrons with a background
density distribution.
 \\
In the case of silicon, for example, one of the most important collision mechanisms are electron-phonon 
scatterings due to lattice vibrations of the cry\-stal, which are modeled by  acoustic (assumed elastic) and optical (non-elastic) 
non-polar modes, the latter with a single frequency $\omega_{p}$, given by: 
\begin{eqnarray}
S(\bk, \bk') & = & (n_{q} + 1) \, K \, \delta(\enka - \enk + \hw) \nonumber
\\
&&
\mbox{} + n_{q} \, K \, \delta(\enka - \enk - \hw) +  K_{0} \, \delta(\enka - \enk) \, ,
\label{Skkarrow}
\end{eqnarray}
with $K$, $K_{0}$ constants for silicon. 

The symbol $\delta$ indicates the usual Dirac delta distribution
corresponding to the well known Fermi's Golden Rule \cite{book:Lundstrom}.
The constant $n_q$ is related to the phonon occupation factor:
$$
\nq = \left[ \exp \left( \frac{\hw}{\kT} \right) - 1 \right] ^{-1},
$$
where $\kB$ is the Boltzmann constant and $T_L = 300 K$ is the constant lattice temperature. In the Boltzmann equation, the energy band function $\varepsilon (\bk)$ are often taken by a simple band model, e.g. the Kane band model 
\eqref{kaneband} or the parabolic band model. Since this term appears in both the transport and collision part, it is evident that it plays an important role in the numerical simulation. We will discuss more about the treatment of this term in later sections.

The self consistent field $ \bE (\ti,\bx)$ is solved from the Poisson equation 
$$
\nabla_{\bx} \cdot \left[ \epsilon_{r}(\bx) \, \nabla_{\bx} V(\ti,\bx) \right]
 = \frac{q}{\epsilon_{0}} \left[ \rho(\ti,\bx) - N_{D}(\bx) \right] ,
\quad
\bE (\ti,\bx) = - \nabla_{\bx} V (\ti,\bx) \, 
$$
where the parameter $\epsilon_{0}$ is the dielectric constant in a vacuum, 
$\epsilon_{r}(\bx)$ labels the relative dielectric function which depends on the material. $\rho(\ti,\bx)$, the electron charge density, is given by the integral over the domain in 
the $\bk$-space $\domk$:
\begin{equation}
\rho(\ti,\bx) = \intd f(\ti, \bx, \bk) \, d \bk \, ,
\end{equation}
and $N_D(\bx)$ is the doping profile, representing an external fixed density of positive charge carriers. 
The evolution is subject to the initial condition
$$
 f(0,\bx,\bk) = f_0(\bx,\bk) \quad \forall \, (\bx,\bk) \in \Omega_{\bx} \times \domk, 
 \, t=0, 
$$
and suitable boundary conditions for $f$ on $\partial \Omega_{\bx} \times \domk$ and 
$\domk \times \partial \Omega_{\bx}$, and for $V$ on $\partial \Omega_{\bx}$ are satisfied. The boundary $\partial \Omega_{\bx} $
is usually split for the Poisson Equation in 
Dirichlet $\partial \Omega^D_{\bx} $, Neumann $\partial \Omega^N_{\bx}$, and Interface boundaries $\partial \Omega^I_{\bx} $,
such that  
$\partial \Omega_{\bx} = \partial \Omega^D_{\bx} \cup \partial \Omega^N_{\bx} \cup  \partial \Omega^I_{\bx} $ .

Examples of boundary conditions used for the Boltzmann Eq. include 
\cite{ref:MarkowichRS}:

\begin{itemize}
\item 
Charge neutrality 
\cite{ref:CerciGambaLev}, \cite{cgms06}, 
\cite{Cheng_08_JCE_BP}, \cite{CGMS-CMAME2008}:

\begin{equation}
f_{out}(\ti,\bx,\bk) = 
\dfrac{N_D(\bx) \, f_{in}(\ti,\bx,\bk)}{\rho_{in}(\ti,\bx)}
\qquad \ti \geq 0, \, x \in \partial \Omega^D_{\bx}, \, \bk \in \domk
\end{equation}
This condition is usually employed at the device contacts
(Dirichlet boundaries $\partial \Omega^D_{\bx} $). 
\item
Null $\bx$-flux:
\begin{equation}
\mathbf{n}(\bx) \cdot \nabla_{\bx} f(\ti,\bx,\bk) = 0 
\qquad \ti \geq 0, x \in \partial \Omega^N_{\bx}, \, \bk \in \domk,
\end{equation}
where $\mathbf{n}(\bx)$ is the normal to the surface $\partial \Omega^N_{\bx}$ at the point $\bx$.
This condition is imposed on the part of the physical domain with an
insulating layer (Neumann boundaries $\partial \Omega^N_{\bx}$).
\item
Vanishing boundary conditions in the $\bk$-space:
\begin{equation}
f(\ti,\bx,\bk) = 0
\qquad \ti \geq 0, \, \bx \in \Omega_{\bx}, \, \bk \in \partial \Omega_{\bk}.
\end{equation}
These conditions correspond to negligible densities 
for large energy values. We use these vanishing conditions for the Boltzmann Equation in our work.
We will just mention that, if we had chosen $\domk$ as the first Brillouin zone, then periodic boundary conditions in the $\bk$-space would be the correct physical conditions. 
However, 
it is difficult to apply these conditions on the complex shape of the boundary 
of a truncated octahedron, which is the shape of the first Brillouin zone for Silicon and Germanium crystals.

Boundary conditions related to the Poisson Equation could be:

\item
Applied potential (bias):
\begin{equation}
V(\ti,\bx)  =  V_{0}(\ti,\bx) 
\qquad \ti \geq 0, \, \bx \in \partial \Omega^D_{\bx} 
\end{equation}
This condition is imposed where we have device contacts (Dirichlet boundaries).
\item
Neumann boundary conditions for the electric potential:
\begin{equation}
\mathbf{n}(\bx) \cdot \nabla_{\bx} V(\ti,\bx) = 0 
\qquad \ti \geq 0 \, , \bx \in \partial \Omega^N_{\bx}
\end{equation}
where $\mathbf{n}(\bx)$ is the normal to the surface $\partial \Omega^N_{\bx}$ at the point $\bx$.
This condition is imposed on the part of the physical domain with an
insulating layer, which is a Neumann boundary.

It is important to mention that the contact boundaries $\partial \Omega^D_{\bx}$ for the Boltzmann and Poisson equations must be the same.

\end{itemize}
\section{Boltzmann equation in spherical coordinates for the $\bk$-vector}
We show here the Boltzmann equation with the momentum $\bk$ in spherical coordinates presented in 
\cite{CGMS-IWCE13}. As opposed to the previous work in \cite{CGMS-CMAME2008}, the coordinate 
transformation 
based on the Kane 
analytic band relation
proposed in \cite{MP} can no longer be used for an energy band that does not assume this analytic Kane band model and that takes into account anisotropy for $\varepsilon(\bk)$. The spherical 
coordinate system is used in
$\bk$ space instead of Cartesian coordinates because of the 
higher resolution demands near the conduction band minimum (chosen as the origin $\bk=0$), and 
large cells in $\bk$-space are sufficient for describing the tail of the distribution 
function accurately.

The following change of variables into 
dimensionless quantities are introduced for a general problem:
\\[5pt]
$
\dm t = \frac{\ti}{t_*} \, , \quad
(\vec{x},z) = \frac{\bx}{\ell_*} \, , \quad
\bk = \frac{\sqrt{2 m^* \kbt}}{\hbar} \, \sqrt{r} \left( \mu,
\sqrt{1 - \mu^{2}} \cos \varphi,  \sqrt{1 - \mu^{2}} \sin
\varphi\right)
$
\\[5pt]
with $r \geq 0 \, , \mu \in [-1,1] \, , \varphi \in [-\pi, \pi]$.
\\[5pt]
$
\dm
\eps(r,\mu,\varphi) = \frac{1}{\kbt}  \varepsilon (\bk)
$
\\[5pt]
$
\dm \nV(t,x,y,z) = \frac{V(t_{*} \, t, \ell_* \, x, \ell_* \, y, \ell_* \, z)}{V_*} , \,
\quad \bE = - \cd{v} \nabla_{\bx} V  
$
\\[5pt]
with $\dm \cd{v} = \dfrac{V_*}{\ell_* E_*}$ and  $E_* = 0.1 \, V_* \, \ell_{*}^{-1}$.
\\[5pt]
where 
the spherical coordinate transformation maps the $\bk$-domain $\domk$ onto the
set $\Omega$ of the $(r, \mu, \varphi)$ space.
Typical values for length, time and voltage are given by
$\ell_* = 10^{-6} \, m$, $t_* = 10^{-12} \, s$ and $V_* = 1 \,
\mbox{Volt}$, respectively.

Thus, a new unknown ``weighted" \emph{pdf} function $\Phi$ is obtained by multiplying the \emph{pdf} $f$
by the Jacobian of the spherical $\bk$-transformation:
\begin{equation}
\Phi(t,x,y,z,r,\mu,\varphi) = \frac{\sqrt{r}}{2} \, f(t,x,y,z,r,\mu,\varphi) \, ,
\end{equation}
which can be interpreted as the probability density function of an electron 
being in the neighborhood of the phase-space state $(x,y,z,r,\mu,\varphi)$ at time $t$.

Hence, writing the collisional integral in spherical coordinates 
and multiplying the Boltzmann equation by the Jacobian associated to the $\bk$-spherical transformation, yields the following Transformed Boltzmann
Equation (TBE) for the unknown $\Phi$:

\begin{equation}
\frac{\partial \Phi}{\partial t} +
\frac{\partial \mbox{ }}{\partial x} \left( a_{1} \, \Phi \right) +
\frac{\partial \mbox{ }}{\partial y} \left( a_{2} \, \Phi \right) +
\frac{\partial \mbox{ }}{\partial z} \left( a_{3} \, \Phi \right) + 
\frac{\partial \mbox{ }}{\partial r} \left( a_{4} \, \Phi \right) + 
\frac{\partial \mbox{ }}{\partial \mu} \left( a_{5} \, \Phi \right) + 
\frac{\partial \mbox{ }}{\partial \varphi} \left(a_{6} \, \Phi \right) = C(\Phi)
\label{boltztrans}
\end{equation}
with the following transport terms
$\vec{a} = (a_1,a_2,a_3,a_4,a_5,a_6)^{T}$: 
\begin{align*}
a_{1}(\cdot) &  =  \cd{D} \left( 2 \, \sqrt{r} \, \mu \, \frac{\partial \eps}{\partial r}  + 
\frac{1 - \mu^{2}}{\sqrt{r}} \frac{\partial \eps}{\partial \mu} \right) ,
\\
a_{2}(\cdot) & =  \cd{D} \left(  2 \, \sqrt{r} \, \sqrt{1 - \mu^{2}} \,
\cos \varphi \, \frac{\partial \eps}{\partial r} - \frac{\mu \, \sqrt{1 - \mu^{2}} \, \cos
\varphi}{\sqrt{r}} \, \frac{\partial \eps}{\partial \mu} -
\frac{\sin \varphi}{\sqrt{r} \, \sqrt{1 - \mu^{2}}} \,
\frac{\partial \eps}{\partial \varphi} \right) ,
\\
a_{3}(\cdot) & =  \cd{D} \left( 2 \, \sqrt{r} \, \sqrt{1 - \mu^{2}} \,
\sin \varphi \, \frac{\partial \eps}{\partial r} - \frac{\mu \, \sqrt{1 - \mu^{2}} \, 
\sin \varphi}{\sqrt{r}} \, \frac{\partial \eps}{\partial \mu}  +
\frac{\cos \varphi}{\sqrt{r} \, \sqrt{1 - \mu^{2}}} \,
\frac{\partial \eps}{\partial \varphi} \right) ,
\\
a_{4}(\cdot) & =  - 2 \, \cd{E} \, \sqrt{r} \left[  \mu \,
E_{x}(t,x,y,z) + \sqrt{1 - \mu^{2}} \left( \cos \varphi \, E_{y}(t,x,y,z) +
\sin \varphi \, E_{z}(t,x,y,z) \right) \right] , 
\\
a_{5}(\cdot) & =  - \cd{E} \left[  \frac{1 - \mu^{2}}{\sqrt{r}} \,
E_{x}(t,x,y,z)- \frac{\mu \sqrt{1 - \mu^{2}}}{\sqrt{r}} \left( 
\cos \varphi \, E_{y}(t,x,y,z) + \sin \varphi \, E_{z}(t,x,y,z) \right) \right] ,
\\
a_{6}(\cdot) & =  - \cd{E} \, \frac{1}{\sqrt{r} \, \sqrt{1 - \mu^{2}}} 
\left[  \mbox{}- \sin \varphi \, E_{y}(t,x,y,z) + \cos \varphi \, E_{z}(t,x,y,z) \right]   ,
\end{align*}
and the linear collision operator
\begin{eqnarray}
&&
C(\Phi)(t,x,y,z,r,\mu,\varphi) =  \frac{\sqrt{r}}{2}
\, \int_{\Omega} {\cal S}(r', \mu', \varphi', r, \mu, \varphi) \,
\Phi(t,x,y,z,r',\mu',\varphi') \: d r' \, d \mu' d \varphi' 
\nonumber
\\
&&
\mbox{ } \hspace{40pt} - \Phi(t,x,y,z,r,\mu,\varphi)
\int_{\Omega} {\cal S}(r, \mu, \varphi, r', \mu', \varphi') \, \frac{\sqrt{r'}}{2} 
\: d r' \, d \mu' d \varphi' \, ,  
\label{colltrans}
\end{eqnarray}
where the scattering kernel is
\begin{eqnarray*}
&& {\cal S}(r, \mu, \varphi, r', \mu', \varphi') = \cd{+} \, 
\delta( \eps(r',\mu',\varphi') - \eps(r,\mu,\varphi) + \alp) 
\\
&&
\mbox{} + \cd{-} \, \delta( \eps(r',\mu',\varphi') - \eps(r,\mu,\varphi) - \alp) 
+ \cd{0} \, \delta( \eps(r',\mu',\varphi') - \eps(r,\mu,\varphi) ),
\end{eqnarray*}
accounting for acoustic and optical electron-phonon interaction, 
the main scattering mechanisms in silicon. The constants above are defined as
\begin{eqnarray*}
&& 
\cd{D} = \frac{t_*}{\ell_*} \sqrt{\frac{\kbt}{2 \, \mass}} \, ,
\quad \cd{E} = \frac{t_* q E_*}{\sqrt{2 \mass \kbt}} \, ,
\quad \alp = \frac{\hw}{\kbt} \, , 
\\
&&
(\cd{+}, \cd{-}, \cd{0}) = \frac{2 \mass \, t_*}{\hbar^3} \sqrt{2 \,
\mass \, \kbt} \left[ (n_{q} + 1) K, n_{q} K, K_0  \right] .
\end{eqnarray*}
The details of the derivation of the TBE are collected in the appendix.
The dimensionless Poisson equation is
\begin{equation}
\frac{\partial}{\partial x} \left( \epsilon_{r} \frac{\partial \nV}{\partial x} \right)
+ \frac{\partial}{\partial y} \left( \epsilon_{r} \frac{\partial \nV}{\partial y} \right)
+ \frac{\partial}{\partial z} \left( \epsilon_{r} \frac{\partial \nV}{\partial z} \right)
= \cd{p} \left[ \rho(t,x,y,z,t) - \mathcal{N}_{D}(x,y,z) \right]
\label{poistrans} 
\end{equation}
where 
\begin{eqnarray*} 
&&  
\mathcal{N}_{D}(x,y,z) =
\left( \frac{\sqrt{2 \,\mass \kbt }}{\hbar} \right)^{\! \! -3}
N_{D}(\ell_* x, \ell_* y, \ell_* z),
\quad
\cd{p} = \left( \frac{\sqrt{2 \,\mass \kbt }}{\hbar} \right)^{\!\! 3}
\frac{\ell_*^{2} q}{\epsilon_{0} \, V_{*}} \, ,
\\
&& 
\rho(t,x,y,z) =
\int_{\Omega} \Phi(t,x,y,z,r',\mu',\varphi') \: d r' \, d \mu' d \varphi' \, .
\end{eqnarray*}
\subsection{Geometrical interpretation of the force terms in the TBE}
Although the terms $(a_1,a_2,a_3)$ related to the transport in the
$\bx$-space due to the electron group 
velocity in the TBE can be easily interpreted as just the gradient $\nabla_{\bk} 
\enk$ expressed in spherical coordinates, the terms $(a_4,a_5,a_6)$ related to the transport in the 
$\bk$-space due to the electric field might be more obscure to understand.
A simple expression for them can be identified.
\begin{align}
a_{4} \, = \, 
- 2 \, \cd{E} \, \sqrt{r} \left(\mu,\sqrt{1 - \mu^{2}} \cos \varphi, \sqrt{1 - \mu^{2}}
\sin \varphi\right)\cdot \bE &  =   - 2 \, \cd{E} \, \sqrt{r} \, \hat{e}_r\cdot \bE, 
\\[5pt]
a_{5} \, = \, - \cd{E} \frac{\sqrt{1 - \mu^{2}}}{\sqrt{r}} \left( \sqrt{1 - \mu^{2}}, -
\mu \cos \varphi, - \mu \sin \varphi\right) \cdot \bE & =
 - \cd{E} \frac{\sqrt{1 - \mu^{2}}}{\sqrt{r}} \, \hat{e}_{\mu} \cdot \bE, 
\\ 
a_{6} \, = \, - \cd{E} \, \frac{1}{\sqrt{r} \, \sqrt{1 - \mu^{2}}} \left(0, -\sin \varphi, 
\cos \varphi \right) \cdot \bE &  =  - \cd{E} \, \frac{1}{\sqrt{r} \, \sqrt{1 - \mu^{2}}} \, \hat{e}_{\varphi} \cdot \bE. 
\end{align}
These transport terms express 
the acceleration field induced by $\bE$ in spherical coordinates,
as they are related to the negatives of the directional cosines of $\bE$ with respect to the 
unit vectors $\hat{e}_r$, $\hat{e}_{\mu}$, $\hat{e}_{\varphi}$.
Hence, the TBE (\ref{boltztrans}) is written in conservative, divergence form, as a flow in 
the $\bk$-space due to the electric field decomposed in each of the orthogonal components of the 
spherical $\bk$-coordinates. This can be easily derived from the expression for the divergence in
general curvilinear coordinates, applied to the particular case of spherical coordinates
$\bk(r,\mu,\varphi)$. This calculation is included in the Appendix.

\section{Computation of the spherical average of a local EPM conduction 
band for silicon}
%
The motivation of this work is to incorporate numerically, in a DG solver of the BP system
electronic conduction bands whose values are obtained by the radial averaging of the full band 
structure given by a local empirical pseudopotential method (EPM) around a local minimum of the 
conduction band for silicon. By performing the radial averaging, it simplifies the discussion of the numerical scheme significantly. 
This is done as a midpoint between a radial and an 
anisotropic full energy band models, with the goal of providing a more accurate physical description of the 
electron group velocity and of the scattering mechanisms by Fermi Golden Rule, and consequently improve the transport and electron - phonon collision
phenomena. The approximation of the electron group velocity is obtained from the numerical values of the derivatives of the 
conduction band, which are obtained by means of a cubic spline interpolation. 
The numerical values of the spherically averaged EPM 
band and the derivatives are obtained as described below.

A local empirical pseudopotential method (EPM) code developed by Chelikowsky et al. 
\cite{paperCHLKW} is adapted to compute the conduction band structure of 
silicon in its Brillouin 
Zone in the $\bk$-space. The local pseudopotentials are used in this EPM code to mimic a silicon semiconductor with crystal diamond structure \cite{COHEN-CHLKW}.

\begin{figure}
\includegraphics[ height=10cm, width=9cm]{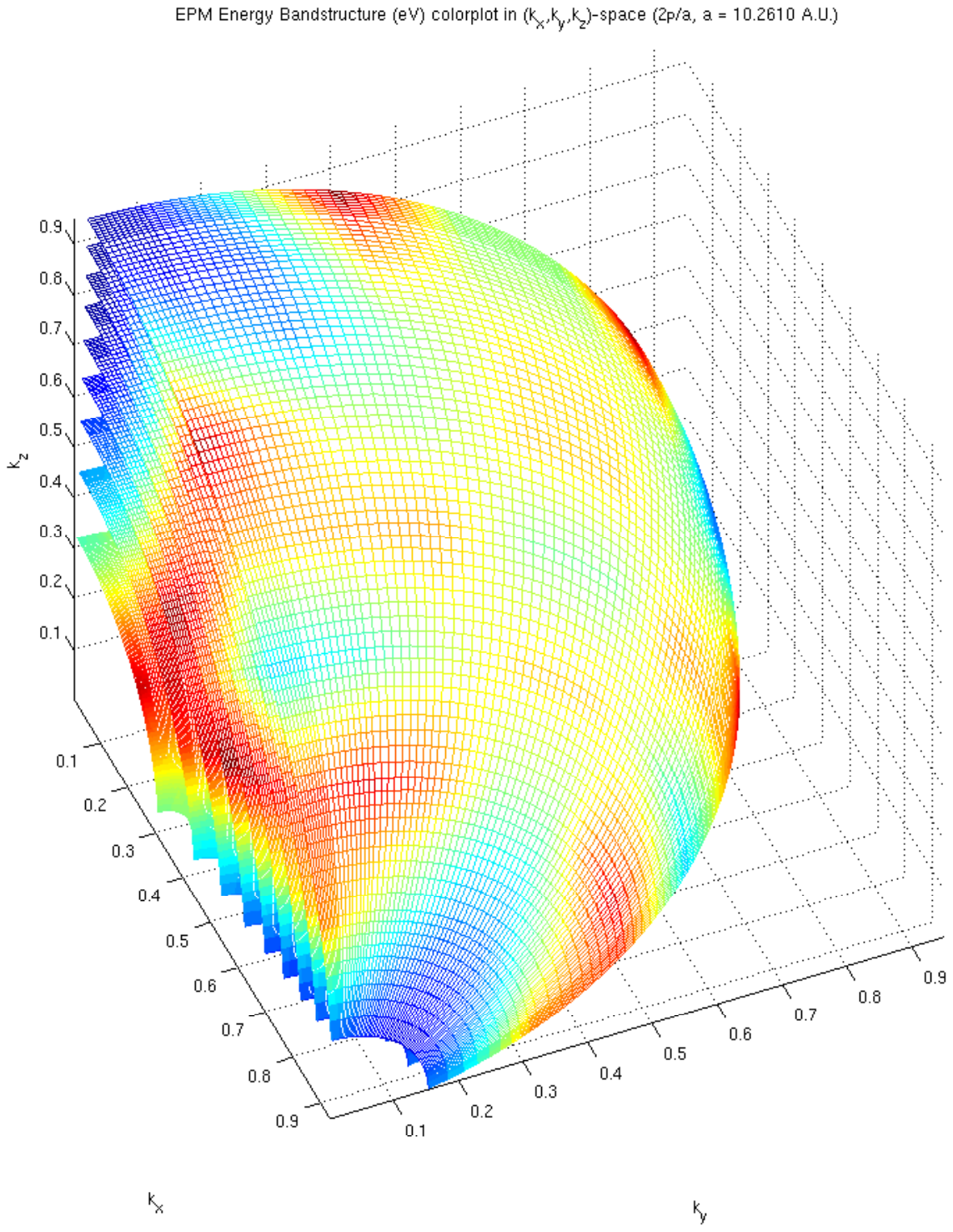}
\caption{Local EPM conduction energy band-structure ($\varepsilon$) color 
plot in the $\bk$-space 
$1^{st}$ octant enclosing the 
Silicon Brillouin Zone. Conduction band local minimum: 
$k_o = (0.8562,0,0) (2\pi/a)$}
\label{EvsK3D}
\end{figure}

A color plot of the local EPM conduction band 
on the first octant of the 
$k$-space enclosing the Brillouin Zone for silicon is shown in Fig. \ref{EvsK3D}. 

The calculated EPM band structure
$\varepsilon(k_x,k_y,k_z) = \varepsilon(\bk(r,\mu,\varphi))$ is then 
averaged over the $\bk$-spheres 
$r_k$ around the local energy minimum point $\bk_0=(0.8562,0,0)2\pi/a$ (where $a$ is the lattice 
constant for silicon) by means of a 10 point Gaussian quadrature on the angular space. Using the 
symmetry of the 
silicon conduction band, the integration only needs to be performed in the 
$(\mu,\varphi)$ domain $[0,1]\times[0,\pi]$ 
\begin{equation}
\tilde{\varepsilon}(r_k) = 
\dfrac{ \dm
\int_{0}^{1} \int_{0}^{\pi} \varepsilon(r_k,\mu,\varphi) \, d\mu d\varphi }{
\dm \int_{0}^{1} \int_{0}^{\pi} d\mu \, d\varphi}
\approx 
\sum_{m=1}^{10} \sum_{n=1}^{10} \omega_{m} \, \omega_{n} \,
\varepsilon(r_k,\mu_m,\varphi_n) \, .
\end{equation}
\ \\

The values of the radius of these $\bk$-spheres are the grid points $r_k$ in the DG-BP simulations. 
In this way we obtain a band model that has a dependence on $r$, and at the same time it uses the
information of the anisotropic energy band values in the angular $\bk$-domain via its numerical average. 
As a midpoint between a radial band model and a full band anisotropic model, it has the desired advantages of both. 
\begin{figure}
\includegraphics[scale=0.6]{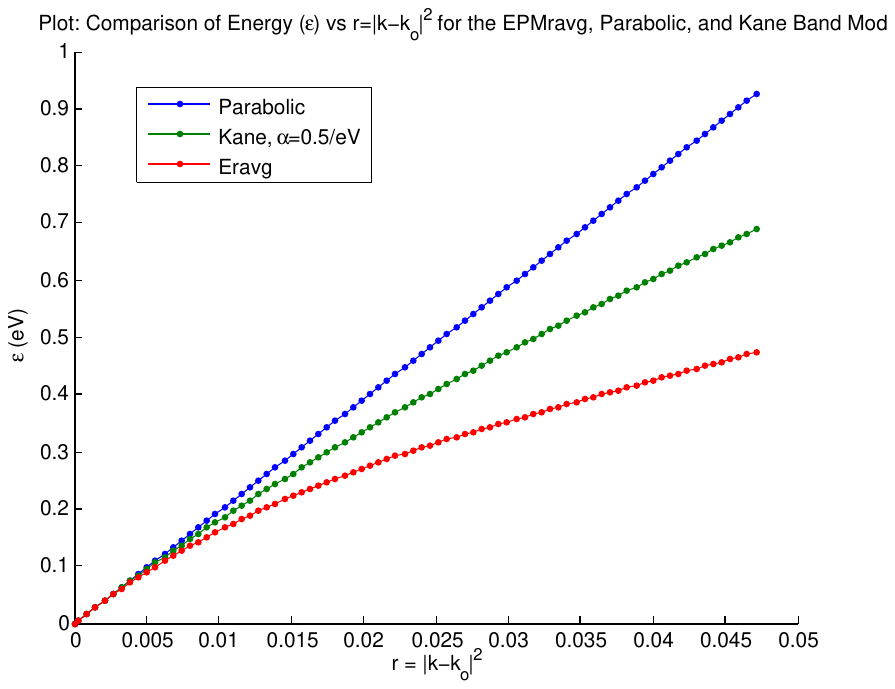}
\caption{Energy (E) vs $r = |\bk-\bk_0|^2$ for the parabolic, Kane, \& 
symmetrized EPM 
spherical average band.}
\label{fig_EnergyParabKaneEPMvsK}
\end{figure}
A cubic spline interpolation is then performed, using the numerical values of the radial average 
$\tilde{\varepsilon}(r) $ at the midpoints of the $r$-cells, and the derivative of this spline 
interpolation is used to obtain a numerical approximation of the derivative ${d 
\tilde{\varepsilon}}/{d r}$ at these $r$-midpoints. 

The spherical averages of the EPM conduction band $\tilde{\varepsilon}(r)$ vs $r \propto 
|\bk-\bk_0|^2$ with the related spline interpolation for Si are shown in Fig. 
\ref{fig_EnergyParabKaneEPMvsK} (in red). The parabolic (blue), which is a linear function of $r$, 
and the Kane (green) analytic conduction band models for silicon are plotted as well. 

It can be observed that 
there is a quantitative difference between the different energy band models. The spherical 
average of the EPM band is below the Kane band model, which is below the Parabolic band. \\

\begin{figure}
\includegraphics[scale=0.6]{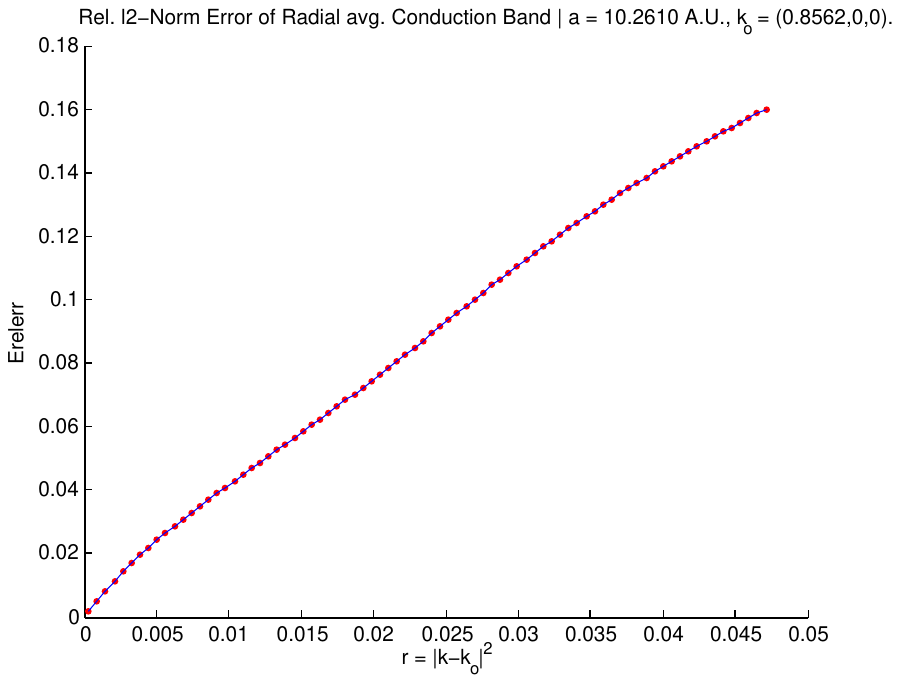}
\caption{Relative $l_2$-deviation in the angular 
space of EPM average from the EPM full band for Si, as a function of $r = |\bk-\bk_0|^2$}
\label{fig_L2errEPMvsr}
\end{figure}

We show in Fig. \ref{fig_L2errEPMvsr}  the relative $l_2$ error norm of the spherical 
average EPM band with respect to the local EPM data $\varepsilon(r,\mu,\varphi)$ as a function of $r$, given by the formula

$$
\frac{\left\langle[\varepsilon-\tilde{\varepsilon}]^2\right\rangle}{\left\langle \varepsilon^2 
\right\rangle}(r_k)
\approx
\frac{
\sum_{m=1}^{10} \sum_{n=1}^{10} 
\omega_{m} \omega_{n}
\left[\varepsilon(r_k,\mu_m,\varphi_n)-\tilde{\varepsilon}(r_k)\right]^2}{
\sum_{m=1}^{10} \sum_{n=1}^{10} 
\omega_{m} \omega_{n}
\left[ \varepsilon(r_k,\mu_m,\varphi_n) \right]^2 \, .
}
$$
\ \\
It can be observed in Fig. \ref{fig_L2errEPMvsr} that the relative $l_2$ error increases with $r$, 
which indicates that far away from the local minimum $\bk_0$ the anisotropy of the conduction band 
becomes increasingly more important. 
%
%

%

\section{DG formulation for the TBE and the Poisson equation} 
%
%
%
In this section, we will discuss the DG schemes based on the radially symmetric band. The scheme is implemented based on a piecewise linear polynomial approximation.

\subsection{Domain and Finite Element Space}
Let's consider a $2 \, D$ rectangular domain in the physical space and a rectangular domain $\domk$ in momentum space.
We use simple rectangular cells
$$
 \Omega_{ijkmn} = \left[ x_{i - \ot} , \, x_{i + \ot} \right] \times
                  \left[ y_{j - \ot} , \, y_{j + \ot} \right] \times
K_{kmn}
$$
where
$$
K_{kmn} =
          \left[ r_{k - \ot} , \, r_{k + \ot} \right] \times
          \left[ \mu_{m - \ot} , \, \mu_{m + \ot} \right] \times
          \left[ \varphi_{n - \ot} , \, \varphi_{n + \ot} \right] 
$$
with
$$
 x_{i \pm \ot} = x_{i} \pm \frac{\Delta x_{i}}{2}, \,
 y_{j \pm \ot} = y_{j} \pm \frac{\Delta y_{j}}{2}, \,
 r_{k \pm \ot} = r_{k} \pm \frac{\Delta r_{k}}{2} \cdots \, ,
$$
and $i=1,...,N_x, \, j=1,..., N_y, \, k=1,...N_r, \, m=1,...N_{\mu}, \, n=1,....,N_\varphi$.

The test functions $\psi(x,y,r,\mu,\varphi)$ belong to the linear function space
$$
V_h^1 = \left\{ v : v|_{\Omega_{ijkmn}} \in P^1(\Omega_{ijkmn}) \right\} ,
$$
where 
$P^1(\Omega_{ijkmn})$ is the set of polynomials of degree at most $1$ on the cell $ \Omega_{ijkmn}$.

A set of piecewise linear basis functions for $V_h^1$ in the open cell $\mathring{\Omega}_{ijkmn}$ is given by
\begin{equation}
\left\lbrace 
1, \, 2 \, \frac{(x - x_{i})}{\Delta x_{i}} , \, 2 \, \frac{(y - y_{j})}{\Delta y_{j}} ,
\, 2 \, \frac{(r - r_{k})}{\Delta r_{k}} , \, 2 \, \frac{(\mu - \mu_{m})}{\Delta \mu_{m}} ,
\, 2 \, \frac{(\varphi - \varphi_{n})}{\Delta \varphi_{n}} 
\right\rbrace  \label{basic_func}
\end{equation}

Hence, in the cell $\mathring{\Omega}_{ijkmn}$, we approximate our weighted-\emph{pdf} $\Phi$ by a  
piecewise polynomial $\Phi_h$ of first degree in $V^1_h$: 
\begin{eqnarray}
&&
\Phi_h (t,x,y,r,\mu,\varphi) = T_{ijkmn}(t) +  
X_{ijkmn}(t) \, \frac{2(x - x_{i})}{\Delta x_{i}} + 
Y_{ijkmn}(t) \, \frac{2(y - y_{j})}{\Delta y_{j}}  \nonumber
\\
&& 
\mbox{ } +
R_{ijkmn}(t) \, \frac{2(r - r_{k})}{\Delta r_{k}} +
M_{ijkmn}(t) \, \frac{2(\mu - \mu_{m})}{\Delta \mu_{m}} +
P_{ijkmn}(t) \, \frac{2(\varphi - \varphi_{n})}{\Delta \varphi_{n}} \, .
\end{eqnarray}
The charge density 
on $\left[ x_{i - \ot} , \, x_{i + \ot} \right] \times
                  \left[ y_{j - \ot} , \, y_{j + \ot} \right]$
under this piecewise linear approximation   is
\begin{equation}
\rho_h(t,x,y)=\sum_{k=1}^{N_{r}} \sum_{m=1}^{N_{\mu}} \sum_{n=1}^{N_{\varphi}}
\left[ T_{ijkmn}(t) +   X_{ijkmn}(t) \, \frac{2(x - x_{i})}{\Delta x_{i}} 
+ Y_{ijkmn}(t) \, \frac{2(y - y_{j})}{\Delta y_{j}} 
\right] \Delta r_{k} \Delta \mu_{m} \, \Delta \varphi_{n} \, .
\label{dg-density} 
\end{equation}
The problem is then reduced to find, by means of our numerical scheme, the unknowns: 

\begin{equation}
T_{ijkmn}(t), \,   
X_{ijkmn}(t), \, 
Y_{ijkmn}(t), \, 
R_{ijkmn}(t), \, 
M_{ijkmn}(t), \, 
P_{ijkmn}(t). \label{unknowns}
\end{equation}

\subsection{Discontinuous Galerkin Formulation for the TBE}
\label{sec:weakform}

The corresponding weak DG formulation and its corresponding approximation consists on finding
$\Phi_h \in V_h^1$, 
such that for any test function $v_h \in V_h^1$ and a generic cell $K$ of the decomposition of $\Omega_{\bx} \times \domk$, solves:
\begin{eqnarray}
&& 
\int_K \frac{\partial \Phi_h}{\partial t} \, v_h d\sigma - 
\int_K \frac{ \partial v_h }{\partial x} \left( a_{1} \, \Phi_h \right) d\sigma -
\int_K \frac{\partial v_h}{\partial y} \left( a_{2} \, \Phi_h \right) d\sigma  - 
\nonumber 
\\[7pt]
&&
\int_K \frac{\partial v_h}{\partial r} \left( a_{4} \, \Phi_h \right) d\sigma - 
\int_K \frac{\partial v_h}{\partial \mu} \left( a_{5} \, \Phi_h \right) d\sigma -
\int_K \frac{\partial v_h}{\partial \varphi} \left( a_{6} \, \Phi_h \right) d\sigma +
\nonumber \\ 
&& 
F_{x}^{+} - F_{x}^{-} + F_{y}^{+} - F_{y}^{-} + 
F_{r}^{+} - F_{r}^{-} + F_{\mu}^{+} - F_{\mu}^{-} + F_{\varphi}^{+} - F_{\varphi}^{-} = 
\int_K C(\Phi_h) v_h \: d\sigma  , \quad 
\label{boltztransDG}
\end{eqnarray}
where $v_h$ is a test function in $V_h^1$,  
$d \sigma = dx \, dy \, dr \, d\mu \, d\varphi$, and
where the $F^{\pm}$'s terms are boundary integrals over 
four-dimensional boundary surfaces associated to each 5-dimensional volume element $ \Omega_{ijkmn}$, that is:
\begin{eqnarray}
F_{x}^{\pm} = 
\int_{y_{j-\frac{1}{2}}}^{y_{j+\frac{1}{2}}} 
\int_{r_{k-\frac{1}{2}}}^{r_{k+\frac{1}{2}}}
\int_{\mu_{m-\frac{1}{2}}}^{\mu_{m+\frac{1}{2}}} 
\int_{\varphi_{n-\frac{1}{2}}}^{\varphi_{n+\frac{1}{2}}}
\left.
a_1 \hat{\Phi}_h v_h^{\mp} \right|_{x_{i \pm \frac{1}{2}}}
dy dr d\mu d\varphi
\nonumber\\
F_{y}^{\pm} =
\int_{x_{i-\frac{1}{2}}}^{x_{i+\frac{1}{2}}} 
\int_{r_{k-\frac{1}{2}}}^{r_{k+\frac{1}{2}}}
\int_{\mu_{m-\frac{1}{2}}}^{\mu_{m+\frac{1}{2}}} 
\int_{\varphi_{n-\frac{1}{2}}}^{\varphi_{n+\frac{1}{2}}}
\left.
a_2 \hat{\Phi}_h v_h^{\mp} \right|_{y_{j \pm \frac{1}{2}}}
dx dr d\mu d\varphi
\nonumber \\
F_{r}^{\pm} =
\int_{x_{i-\frac{1}{2}}}^{x_{i+\frac{1}{2}}}
\int_{y_{j-\frac{1}{2}}}^{y_{j+\frac{1}{2}}} 
\int_{\mu_{m-\frac{1}{2}}}^{\mu_{m+\frac{1}{2}}} 
\int_{\varphi_{n-\frac{1}{2}}}^{\varphi_{n+\frac{1}{2}}}
\left.
a_4 \hat{\Phi}_h v_h^{\mp} \right|_{r_{k \pm \frac{1}{2}}}
dx dy d\mu d\varphi
\nonumber \\
F_{\mu}^{\pm} = 
\int_{x_{i-\frac{1}{2}}}^{x_{i+\frac{1}{2}}}
\int_{y_{j-\frac{1}{2}}}^{y_{j+\frac{1}{2}}} 
\int_{r_{k-\frac{1}{2}}}^{r_{k+\frac{1}{2}}}
\int_{\varphi_{n-\frac{1}{2}}}^{\varphi_{n+\frac{1}{2}}}
\left.
a_5 \hat{\Phi}_h v_h^{\mp} \right|_{\mu_{m \pm \frac{1}{2}}}
dx dy dr d\varphi
\nonumber \\
F_{\varphi}^{\pm} =
\int_{x_{i-\frac{1}{2}}}^{x_{i+\frac{1}{2}}}
\int_{y_{j-\frac{1}{2}}}^{y_{j+\frac{1}{2}}} \int_{r_{k-\frac{1}{2}}}^{r_{k+\frac{1}{2}}}
\int_{\mu_{m-\frac{1}{2}}}^{\mu_{m+\frac{1}{2}}}
\left. 
a_6 \hat{\Phi}_h v_h^{\mp} \right|_{\varphi_{n \pm \frac{1}{2}}}
dx dy dr d\mu \quad .
\nonumber
\end{eqnarray}
The values for $v_h^{\mp}$ are the ones for the function $v_h$ on the interior of the considered cell. The upwind numerical flux 
$\hat{\Phi}_h$  defines the value of $\Phi_h$ at the 
boundary. That means $\Phi_h$ might be discontinuous at the boundary.

The collisional terms 
$$\int_K C(\Phi_h) v_h \: d\sigma$$ 
become a linear combination, with numerical constant coefficients, of the the unknowns (\ref{unknowns}) 
which are precomputed and stored. 
The Poisson equation can be solved by either an integral formula, projecting the solution to the electric field into the space $V_h^1$, for the 1D device case, or by means of a LDG method, for higher dimensional cases.
 A Runge  Kutta method is applied for the time evolution of the 
time dependent coefficients (\ref{unknowns}) for the piecewise linear approximation $\Phi_h \in V_h^1$.

\subsection{Calculation of the Collision terms in the DG formulation}

Here, we will show the details of the calculation of the collisional integrals since it is the most demanding part of the simulation. For completeness, the calculation of the transport term is also reported, and collected in the Appendix \ref{sec:trans}.


Denote by 
$
 K_{kmn} = \left[ r_{k - \ot} , \, r_{k + \ot} \right] \times
           \left[ \mu_{m - \ot} , \, \mu_{m + \ot} \right] \times
           \left[ \varphi_{n - \ot} , \, \varphi_{n + \ot} \right]
$ the rectangular cells in the spherical coordinates for $\bk$-space.
Because the collisional operators only perform integrations in $\bk$-space, it is convenient to write the basis functions in (\ref{basic_func})
as the product of two functions $\eta^{p}_{i,j}(x,y)$ and $\xi^{p}_{k,m,n}(r,\mu,\varphi)$, which 
are given in $\Omega_{ijkmn} = \Omega_{I}$ by
\begin{eqnarray}
\left\lbrace \eta^{p}_{i,j}(\vec{x}) \right\rbrace_{p=0,1,..,5} & = &
\left\lbrace 
1 , \, 1 , \, 1 , \, 1 \, ,
\frac{2(x - x_{i})}{\Delta x_{i}} , \, \frac{2(y - y_{j})}{\Delta y_{j}}
\right\rbrace ,
\\[10pt]
\left\lbrace \xi^{p}_{k,m,n}(\vec{r}\,) \right\rbrace_{p=0,1,..,5} & = &
\left\lbrace 
1 , \, \frac{2(r - r_{k})}{\Delta r_{k}} , \, \frac{2(\mu - \mu_{m})}{\Delta \mu_{m}} ,
\, \frac{2(\varphi - \varphi_{n})}{\Delta \varphi_{n}} , \, 1 , \, 1
\right\rbrace \, ,
\end{eqnarray}
where we define
\begin{equation}
\vec{x} = (x,y) \, , \quad \vec{r} = (r,\mu, \varphi) \, , \quad  \vec{r} \, ' = (r',\mu', \varphi') \, , \quad  d\,\vec{r} = dr\,d\mu\,d\varphi
\end{equation}
\begin{equation}
 I  =  (i,j,k,m,n)
\end{equation}
\begin{equation} \label{eq:chiI}
\chi_{I} = \chi_{I} ( \vec{x},\vec{r} \, ) =  \left\lbrace 
\begin{array}{ll}
1 & \mbox{ if } (\vec{x},\vec{r}\,) \in \mathring{\Omega}_{I}  \\
0 & \mbox{ otherwise}
\end{array}
\right. 
\end{equation}
and 
\begin{equation}
 W_{I}^0(t) := T_{I}(t) \, , \quad W_{I}^1(t) := R_{I}(t) \, , \quad W_{I}^2(t) := M_{I}(t) \, , \quad  W_{I}^3(t) := P_{I}(t) \, , 
\end{equation} 
\begin{equation} \label{eq:WpI}
W_{I}^4(t) := X_{I}(t) \, , \quad W_{I}^5(t) := Y_{I}(t) \, . 
\end{equation}

Then, in a piecewise continuous linear approximation of $\Phi$, we have (almost everywhere), that
\begin{equation}
\Phi(t,\vec{x},\vec{r}\,) =
\sum_{I} \chi_{I} ( \vec{x},\vec{r} \, )  
\left[ \, 
\sum_{p=0}^{5} W_{I}^p(t) \, \eta^{p}_{i,j}(\vec{x}) \, \xi^{p}_{k,m,n}(\vec{r} \, ) 
\, \right]
\label{Phi_W}
\end{equation}

Because the phonon collision scatterings 
only consider the Fermi Golden Rule \cite{book:Lundstrom} 
and the spherical coordinates localize the negative part operator,
there is a natural split of the collision operator in gain and loss terms of probability density rates.  

\smallskip 

{\bf Gain Term of the collisional operator.} The {\bf gain term}, when
using the piecewise linear function (\ref{Phi_W}), becomes

{\small
\begin{eqnarray*}
\frac{\sqrt{r}}{2} 
\int_{0}^{\infty} \hspace{-5pt}  \int_{-1}^{1} \hspace{-2pt}
\int_{-\pi}^{\pi} \hspace{-5pt} d\vec{r} \, ' \,
{\cal S}(\vec{r}\,',\vec{r}\,)  
\sum_{I} \chi_{I}(\vec{x} , \vec{r}\, ' )
\sum_{p=0}^{5} W_{I}^p (t) \eta^{p}_{i,j}(\vec{x}) \xi^{p}_{k,m,n}(\vec{r}\,')  
& \approx &
\\
\sum_{I}  \frac{\sqrt{r}}{2} \, \int_{K_{k m n}}
{\cal S}(\vec{r}\,',\vec{r}\,) 
\, \chi_{I}(\vec{x},\vec{r}\,')
\sum_{p=0}^{5} W_{I}^p (t) \, \eta^{p}_{i,j}(\vec{x}) \, \xi^{p}_{k,m,n}(\vec{r}\,') 
d\vec{r} \, ' \,
& = &
\\
\sum_{I} \sum_{p=0}^{5}  \frac{\sqrt{r}}{2} \,
\chi_{ij}(\vec{x}) \, W_{I}^p (t) \, \eta^{p}_{i,j}(\vec{x})
\int_{K_{k m n}}
{\cal S}(\vec{r}\,',\vec{r}\,) \, \xi^{p}_{k,m,n}(\vec{r}\,')
\: d\vec{r} \, ' \,
&&
\end{eqnarray*}
}
with $\chi_I$ and $W^p_I$ from (\ref{eq:chiI} - \ref{eq:WpI})
$$
\chi_{ij}(\vec{x}) = \left\lbrace 
\begin{array}{ll}
 1 & \mbox{ if } (x,y) \in \left[ x_{i - \ot} , \, x_{i + \ot} \right] \times
                  \left[ y_{j - \ot} , \, y_{j + \ot} \right] 
\\
0 & \mbox{ otherwise }
\end{array}
\right. .
$$
%
In the weak formulation, the gain term is multiplied by the test function
$\eta^{q}_{\ib,\jb}(\vec{x}) \, \xi^{q}_{\kb,\mb,\nb}(\vec{r})$
and an integral over the domain $ \Omega_{{\bar{I}}} \, , \, \bar{I}  = ( {\ib , \jb , \kb , \mb , \nb} ) ,$ with respect to $(\vec{x},\vec{r})$ is performed, obtaining:

{\small
\begin{equation}
 \int_{\Omega_{{{ \bar{I} }}}} 
\sum_{I} \sum_{p=0}^{5}  \frac{\sqrt{r}}{2}  \,
\chi_{ij}(\vec{x}) \, W_{I}^p (t) \, \eta^{p}_{i,j}(\vec{x})
\int_{K_{k m n}}
{\cal S}(\vec{r}\,',\vec{r}\,) \, \xi^{p}_{k,m,n}(\vec{r}\,'\,)
\: d\vec{r} \, ' 
\quad
\eta^{q}_{\ib,\jb}(\vec{x}) \, \xi^{q}_{\kb,\mb,\nb}(\vec{r}) \:
d\vec{x} \, d\vec{r}  \,  \label{int_gain}
\end{equation}
or
\begin{equation}
\sum_{I} \sum_{p=0}^{5} W_{I}^p (t) 
\int_{K_{\kb \mb \nb}} \frac{\sqrt{r}}{2} 
\int_{K_{k m n}}
{\cal S}(\vec{r}\,',\vec{r}\,) \xi^{p}_{k,m,n}(\vec{r}\,'\,) \: 
d\vec{r}\,'
\xi^{q}_{\kb,\mb,\nb}(\vec{r}) \: d\vec{r} 
\,
\int_{x_{\ib - \ot}}^{x_{\ib + \ot}} 
\int_{y_{\jb - \ot}}^{y_{\jb + \ot}} 
\chi_{ij}
\, 
\eta^{p}_{i,j}(\vec{x}) \, \eta^{q}_{\ib,\jb}(\vec{x}) \: d\vec{x} \, 
\label{weak_gain}
\end{equation}
}
The integration with respect to $x$ and $y$ gives
$$
\int_{x_{\ib - \ot}}^{x_{\ib + \ot}} 
\int_{y_{\jb - \ot}}^{y_{\jb + \ot}} 
\chi_{ij}(\vec{x}) \, \eta^{p}_{i,j}(\vec{x}) \, \eta^{q}_{\ib,\jb}(\vec{x}) \: dx \, dy  =
\delta_{i \ib} \, \delta_{j \jb} \,
\beta_{p q} \, \Delta x_{\ib} \, \Delta y_{\jb}
$$
where the matrix $\beta_{p q}$ has the following terms:
$$
\left( \beta_{p q} \right) = \left( 
\begin{array}{cccccc}
1 & 1 & 1 & 1 & 0 & 0 \\
1 & 1 & 1 & 1 & 0 & 0 \\
1 & 1 & 1 & 1 & 0 & 0 \\
1 & 1 & 1 & 1 & 0 & 0 \\
0 & 0 & 0 & 0 & \frac{1}{3} & 0 \\ 
0 & 0 & 0 & 0 & 0 & \frac{1}{3}
\end{array}
\right) .
$$
then equation (\ref{weak_gain}) is reduced to:
{
\begin{equation} \label{wxy_gain}
\sum_{k,m,n} \sum_{p=0}^{5} 
W_{\ib \jb kmn}^p(t) \beta_{p q} \Delta x_{\ib} \Delta y_{\jb} \times 
\int_{K_{\kb \mb \nb}} \left[ 
 \frac{\sqrt{r}}{2} \int_{K_{k m n}} \!
{\cal S}(\vec{r}\,',\vec{r}\,) \, \xi^{p}_{k,m,n}(\vec{r}\,') \,
d\vec{r}\,'\right] \,
\xi^{q}_{\kb,\mb,\nb}(\vec{r}) \: 
 d\vec{r} .
\end{equation}
}

\smallskip

{\bf Loss Term of the collisional operator.} The weak formulation of the {\bf loss term} of the collisional operator gives
{\small
\begin{eqnarray}
&&
\int_{\Omega_{{\bar{I}}}} 
\Phi(t,\vec{x},\vec{r}) 
\int_{0}^{+ \infty} \hspace{-5pt} dr' \int_{-1}^{1} \hspace{-2pt} d \mu'
\int_{-\pi}^{\pi} \hspace{-5pt} d\varphi' \:
{\cal S}(\vec{r},  \vec{r}\,') \frac{\sqrt{r'}}{2} \,
\eta^{q}_{\ib,\jb}(\vec{x}) \, \xi^{q}_{\kb,\mb,\nb}(\vec{r}) \,
d\vec{x} d\vec{r}  \, \approx
\label{weak_loss}
\\[8pt]
&&
\int_{\Omega_{{\bar{I}}}} 
\Phi(t,\vec{x},\vec{r}) 
\left[ 
\sum_{k,m,n} \int_{K_{k m n}} \frac{\sqrt{r'}}{2}
{\cal S}(\vec{r},  \vec{r}\,')  d\vec{r}\,' 
\right]
\eta^{q}_{\ib,\jb}(\vec{x})  \xi^{q}_{\kb,\mb,\nb}(\vec{r}) \,
d\vec{x} d\vec{r} 
\nonumber
\end{eqnarray}
}
Using the linear approximation of $\Phi$ given by (\ref{Phi_W}), integral (\ref{weak_loss}) becomes
{\small
\begin{eqnarray*}
&&
\int_{\Omega_{{\bar{I}}}} 
\left[ 
\sum_{I} \! \chi_{I}(\vec{x},\vec{r})
\sum_{p=0}^{5} W_{I}^p (t) \, \eta^{p}_{i,j}(\vec{x}) \, \xi^{p}_{k,m,n}(\vec{r})
\right] 
\left[ \sum_{k,m,n} \int_{K_{k m n}} \frac{1}{2} \, \sqrt{r'} \,
{\cal S}(\vec{r},  \vec{r}\,') \: d\vec{r} \, ' \, \right]
\times
\eta^{q}_{\ib,\jb}(\vec{x}) \, \xi^{q}_{\kb,\mb,\nb}(\vec{r}) \:
d\vec{x} \, d\vec{r}  
\, =
\\[8pt]
&&
\sum_{p=0}^{5} W_{{\bar{I}}}^p(t)
\int_{\Omega_{{\bar{I}}}} 
\eta^{p}_{\ib,\jb}(\vec{x})  \xi^{p}_{\kb,\mb,\nb}(\vec{r})
\left[ 
\sum_{k,m,n} \int_{K_{k m n}} \frac{\sqrt{r'}}{2}
{\cal S}(\vec{r},  \vec{r}\,')  d\vec{r}\,' \right]
\eta^{q}_{\ib,\jb}(\vec{x})  \xi^{q}_{\kb,\mb,\nb}(\vec{r}) \,
d\vec{x} d\vec{r} \, .
\end{eqnarray*}
}

Therefore, equation (\ref{weak_loss}) reduces to
{\small
\begin{equation} \label{wxy_loss}
\sum_{p=0}^{5} W_{{\bar{I}}}^p(t)
 \beta_{p q}  \Delta x_{\ib}  \Delta y_{\jb}
\sum_{k,m,n} 
\int_{K_{\kb \mb \nb}} \!
\left[ 
\int_{K_{k m n}} \! 
\frac{\sqrt{r'}}{2} \,
{\cal S}(\vec{r},  \vec{r}\,') \: 
d\vec{r}\,'\right] \!
\xi^{p}_{\kb,\mb,\nb}(\vec{r}) \,
\xi^{q}_{\kb,\mb,\nb}(\vec{r}) \:
d\vec{r}  
\end{equation}
}


In the case of an energy band function with radial dependance $\varepsilon(r)$, 
\begin{eqnarray}
{\cal S}(r, r')  & = & 
\cd{0} \, \delta( \eps(r') - \eps(r) )  +  \cd{+} \, \delta( \eps(r') - \eps(r) + \alp) 
 + \cd{-} \, \delta( \eps(r') - \eps(r) - \alp) 
\nonumber \\
& = & \sum_{l=-1}^{+1} \cd{l} \, \delta( \eps(r') - \eps(r) + l \alp) \, .
\end{eqnarray}

The radial energy band function can be projected on the space of piecewise linear functions of $r$ to obtain

\begin{equation}
\eps_{h}(r) = \sum_{k=1}^{N_r} \chi_k  \left[ \eps(r_k) + A_k (r - r_k) \right] = \sum_{k=1}^{N_r} \chi_k  \left[ \, \eps(r_k) \,  +  \, \partial_{r}\eps(r_k) \, (r - r_k) \, \right] \, ,
 \end{equation}

and, after this projection, we can calculate the collision integrals involving a delta distribution
with the piecewise linear function in their argument. 
The computation of such collisional integrals are given by
\begin{eqnarray}
\int_K C(\Phi_h) v_h \: d\sigma  
& = & 
\sum_{k,m,n} \sum_{p=0}^{5} 
W_{\ib \jb kmn}^p \beta_{p q} \Delta x_{\ib} \Delta y_{\jb}  
\int_{K_{\kb \mb \nb}} 
 \int_{K_{k m n}} \!
{\cal S}(r', r ) \, \xi^{p}_{k,m,n}(\vec{r}\,'\,) \,
d\vec{r}\,'
\frac{\sqrt{r}}{2} \,
\xi^{q}_{\kb,\mb,\nb}(\vec{r}) \: 
 d\vec{r} 
\nonumber\\
& -  &  
\sum_{p=0}^{5} W_{{\bar{I}}}^p(t)
 \beta_{p q}  \Delta x_{\ib}  \Delta y_{\jb}
\sum_{k,m,n} 
\int_{K_{k m n}} \! 
\left[ 
\int_{K_{\kb \mb \nb}} \!
{\cal S}(r, r' ) \: 
\xi^{p}_{\kb,\mb,\nb}(\vec{r}) \,
\xi^{q}_{\kb,\mb,\nb}(\vec{r}) \:
d\vec{r}  
\right] \!
\frac{\sqrt{r'}}{2} \,
d\vec{r}\,'\nonumber \\
& = &  
\sum_{k,m,n} \sum_{p=0}^{5} 
W_{\ib \jb kmn}^p(t) \beta_{p q} \Delta x_{\ib} \Delta y_{\jb}  
\times \\
& &
\int_{K_{\kb \mb \nb}} \left[ 
 \int_{K_{k m n}} \!
\sum_{l=-1}^{+1} \cd{l} \, \delta( \eps(r) - \eps(r') + l \alp) \, \xi^{p}_{k,m,n}(\vec{r}\,'\,) \,
d\vec{r}\,'\right] \,  \frac{\sqrt{r}}{2} \,
\xi^{q}_{\kb,\mb,\nb}(\vec{r}) \: 
 d\vec{r} 
\nonumber\\
& - & 
\sum_{p=0}^{5} W_{{\bar{I}}}^p(t)
 \beta_{p q}  \Delta x_{\ib}  \Delta y_{\jb}
\times \nonumber \\
& & 
\sum_{k,m,n} 
\int_{K_{k m n}} \! 
\left[ 
\int_{K_{\kb \mb \nb}} \!
\sum_{l=-1}^{+1} \cd{l} \, \delta( \eps(r') - \eps(r) + l \alp) \: 
\xi^{p}_{\kb,\mb,\nb}(\vec{r}) \,
\xi^{q}_{\kb,\mb,\nb}(\vec{r}) \:
d\vec{r}  
\right] \! 
\frac{\sqrt{r'}}{2} \,
d\vec{r}\,' \, . \nonumber 
\end{eqnarray}

In order to perform the integrations involving $\sqrt{r}$ numerically by means of Gaussian quadrature,
the change of variables $r = s^2$ is applied, so that the functions of $s$ to be integrated
are just polynomials, 
\begin{eqnarray}
& & \int_K C(\Phi_h) v_h \: d\sigma =  \nonumber\\
& & 
\sum_{k,m,n} \sum_{p=0}^{5} 
W_{\ib \jb kmn}^p(t) \beta_{p q} \Delta x_{\ib} \Delta y_{\jb}  
\int_{K_{\kb \mb \nb}} 
\frac{s}{2}  \,
\xi^{q}_{\kb,\mb,\nb}(s^2,\mu,\varphi) \,  2s  \:
\times 
\nonumber \\
& & 
\left[ 
\sum_{l=-1}^{+1} \cd{l} \int_{K_{k m n}} 
\delta( \eps(r_{\kb}) + A_{\kb}(s^2 - r_{\kb} ) + l \alp - \eps(r_k) - A_k (r' - r_k) ) \, \xi^{p}_{k,m,n}(\vec{r}\,'\,) \,
d\vec{r}\,' \right] ds d\mu d\varphi 
\nonumber\\
&  & - \,
\sum_{p=0}^{5} W_{{\bar{I}}}^p(t)
 \beta_{p q}  \Delta x_{\ib}  \Delta y_{\jb} \,
\sum_{k,m,n} 
\int_{K_{k m n}} \! 
\frac{{s'}}{2} \, 2s'
\times \nonumber \\
& & 
\left[ 
\sum_{l=-1}^{+1} \cd{l}
\int_{K_{\kb \mb \nb}} \!
\delta( \eps(s'^2) + l \alp - \eps(r_{\kb}) - A_{\kb} (r - r_{\kb})  ) \: 
\xi^{p}_{\kb,\mb,\nb}(\vec{r}) \,
\xi^{q}_{\kb,\mb,\nb}(\vec{r}) \:
d\vec{r}  
\right] \! 
ds'  d\mu'  d\varphi' 
\nonumber \\
& &  =
\sum_{k,m,n} \sum_{p=0}^{5} 
W_{\ib \jb kmn}^p(t) \beta_{p q} \Delta x_{\ib} \Delta y_{\jb} \times 
\\
&&   
\int_{K_{\kb \mb \nb}} 
 \,
\xi^{q}_{\kb,\mb,\nb}(s^2,\mu,\varphi)  \:
\chi_{k}\left(\frac{\eps(r_{\kb}) + A_{\kb}(s^2 - r_{\kb} ) + l \alp - \eps(r_k) + A_k r_k }{A_k}\right)
\times 
\nonumber \\
& & 
\left[ 
\sum_{l=-1}^{+1} \cd{l} 
\int_{\mu_{m-\frac{1}{2}}}^{\mu_{m+\frac{1}{2}}} \int_{\varphi_{n-\frac{1}{2}}}^{\varphi_{n+\frac{1}{2}}}
\xi^{p}_{k,m,n}\left( \frac{\eps(r_{\kb}) + A_{\kb}(s^2 - r_{\kb} ) + l \alp - \eps(r_k) + A_k r_k }{A_k} ,\mu',\varphi' \right) d\mu' d\varphi' 
\right] 
{s^2} 
ds d\mu d\varphi 
\nonumber\\
&  & - \,
\sum_{p=0}^{5} W_{{\bar{I}}}^p(t)
 \beta_{p q}  \Delta x_{\ib}  \Delta y_{\jb} \,
\sum_{k,m,n} 
\int_{K_{k m n}} \! 
d\mu'  d\varphi' \,
{{s'}^2} \,
\chi_{\kb}\left( \frac{ \eps(r_k) + A_k(s'^2 - r_k) + l \alp - \eps(r_{\kb}) + A_{\kb} r_{\kb} }{A_{\kb}}\right)
\times
\nonumber \\
& & 
\left.
\left[ 
\sum_{l=-1}^{+1} \cd{l} \: 
\int_{\mu_{\mb - \frac{1}{2}}}^{\mu_{\mb + \frac{1}{2}}} \int_{\varphi_{\nb - \frac{1}{2}}}^{\varphi_{\nb + \frac{1}{2}}}
\xi^{p}_{\kb,\mb,\nb}(r(s),\mu,\varphi) \,
\xi^{q}_{\kb,\mb,\nb}(r(s),\mu,\varphi) \:
d\mu d\varphi  
\right] \! 
\right|_{r(s)} ds'   \, ,
\nonumber 
\end{eqnarray}

with $ r(s) = \frac{  \eps(r_k) + A_k(s'^2 - r_k) + l \alp - \eps(r_{\kb}) + A_{\kb} r_{\kb} }{A_{\kb}} $.

The integrals above involve only polynomials, which are numerically computed by Gaussian quadrature rules.

\subsection{The algorithm for time evolution}
Starting with given initial and boundary conditions, the algorithm advances
from $t^n$ to $t^{n+1}$ by the following way:
\begin{enumerate}
\item 
Compute the density $\rho$.
\item
Solve the Poisson equation and find the electric field $\bE$.
\item
Compute the transport terms $a_i$'s.
\item
Compute the collision part.
\item
Solve the (large) system of ordinary differential equations for 
the coefficients of the linear 
approximation of $\Phi_h$ (which are obtained from the DG formulation), by using a TVD Runge - Kutta scheme.
\item 
Repeat the previous steps as needed.
\end{enumerate}

\begin{algorithm}[H]
\SetAlgoLined
\KwData{Given Initial and Boundary Conditions}
\KwResult{Time Evolution of Probability Density Function}
initialization\;
\While{$t_l<T_{\mbox{max}}$}{
Compute density $\rho(\bx,t_l)$\;
Solve Poisson Eq. to find the electric field $\bE(\bx,t_l)$\;
$t_{l+1} = t_l + \Delta t_{l+1}$\;
Compute the collision terms\;
Compute the transport terms $a_i(\bx,\bk,t_{l+1})$'s\;
Solve the large system of ODEs for 
the coefficients of $\left. \Phi_h \right|_{t_{l+1}}$ (obtained from the DG formulation) by a TVD Runge-Kutta scheme.
Repeat the previous steps if needed.
}
\caption{DG-BP Algorithm}
\end{algorithm}

\section{The $n^{+}$-$n$-$n^{+}$ silicon diode} 
We consider the symmetric case of a 1D $n^{+}$-$n$-$n^{+}$ diode, 
in which the conduction 
band energy function is assumed to be of the form 
$\varepsilon(|\bk|) = \varepsilon(r)$. 
This assumption preserves azimuthal symmetry for the problem if the initial condition is independent of the azimuthal direction $\varphi$. Therefore, under these assumptions the problem has 
azimuthal symmetry in $\bk$ for all times $t \geq 0$, so it suffices to consider $\bk = \bk(r,\mu)$, 
reducing then the dimensionality of the problem to 1-D in  $x$-space and 2-D in $\bk = \bk(r,\mu)$, then the problem reduces to a 3-D plus time. Assuming $\bE$ has 
null $y$ and $z$ components, 
this  symmetric case reduces the TBE to
\begin{equation}
\frac{\partial \Phi}{\partial t} +
\frac{\partial \mbox{ }}{\partial x} \left( a_{1} \, \Phi \right) +
\frac{\partial \mbox{}}{\partial r} \left( a_{4} \, \Phi \right)  + 
\frac{\partial \mbox{ }}{\partial \mu} 
\left( a_{5} \, \Phi \right) 
= C(\Phi),
\end{equation}
where the terms $a_{1}$, $a_{4}$ and $a_{5}$ are now simplified.

The Poisson equation is reduced to
\begin{equation}
\frac{\partial}{\partial x} \left( \epsilon_{r} \frac{\partial \nV}{\partial x} \right)
  = \cd{p} \left[\rho(t,x) - \mathcal{N}_{D}(x)\right] .
\end{equation}
For this case both the potential and electric field
have analytic integral solutions, 
that are easily computed numerically for the piecewise linear approximation of the density 
$\rho_h$. Then, such electric field solution is projected in the $V_h^1$ space of piecewise linear polynomials.
\subsection{Device specifics} We consider first a diode of $1 \, \mu m$ length, with an n-channel of $400 \, nm$ length, doping 
of $5 \times 10^{23} \, m^{-3}$ in the $n^+$ region and $2 \times10^{21} \, m^{-3}$ in the $n$ 
region. 
We also consider a $0.25 \mu m$ diode with a $50 \,  nm$ channel with $n^+$-doping of $5\times10^{24} \, m^{-3}$,
and $n$-doping of $1\times10^{21} \, m^{-3}$.
\subsection{Numerical simulations} 

The space $V_h^1$ of piecewise linear polynomials in $(x,r,\mu)$, with time dependent coefficients,
is used as both the trial and test space in our DG scheme. The input data of the numerical simulations is

\begin{itemize}
\item 
{\bf Computational domain:} 
$ x \in [0,1], \, r \in [0,r_{max}], \, \mu \in [-1,1] $,
where $r_{max}$ is taken in the numerical experiments such that
$ \Phi(t,x,r,\mu) \approx 0$ \mbox{for} $ r \approx r_{max} $ 
(for example, $r_{max}=36$ for $V_{bias} = 0.5$ Volts in the 400nm channel case)
\item
{\bf Initial condition:} 
$\dm \Phi(0, x, r,\mu) = \Pi_h \left\lbrace C N_D(x) \frac{\sqrt{r}}{2} e^{-\varepsilon(r)} \right\rbrace $, 
where $C$ constant is such that $\rho(x,0)$ equals the doping $ N_D(x) $ at $t=0$. The initial condition is projected by $\Pi_h $ into $V_h^1$.
\item
{\bf Boundary conditions:} 
Neutral charges at the endpoints $x_{\frac{1}{2}} = 0$ and $x_{N_x+\frac{1}{2}} = 1$. \\[5pt]
$\dm \Phi(t,0,r,\mu) = N_D(0) \frac{\Phi(t,x_1,r,\mu)}{\rho(t,x_1)}$
and
$\dm \Phi(t,1,r,\mu) = N_D(1) \frac{\Phi(t,x_{N_x},r,\mu)}{\rho(t,x_{N_x})}$.

Cut-off in the $k$-space $\dm \Phi(t,x,r_{max},\mu) = 0$.
\item
{\bf Applied potential - bias:} 
$\dm \Psi(t,0)=0$ and $\dm \Psi(t,1) = V_0$.
\end{itemize}
No boundary conditions are needed on $r=0$, $\mu = \pm 1$.
Upwind fluxes in $r$ and $\mu$ are analytically zero at these boundaries, since they are 
related to points in
$\bk$-space such as the origin and the poles, which are transformed into boundaries when applying the spherical 
change of coordinates.
It is very simple to verify that $a_4 = 0$ at $r = 0$, and $a_5 = 0$ at $\mu = \pm 1$.

\section{Numerical results}
The BP transport along the EPM spherical average energy band is
numerically simulated by means of our DG-BP solver, and compared to simulations where 
the values related to the analytical Parabolic and Kane band models are implemented numerically. 
We compare simulations for two $n^+$-$n$-$n^+$ silicon diodes with different characteristics.
The first one has a length of  $1 \mu m$, an n-channel length of $400 nm$, 
$n^+$ doping of ${5 \cdot 10^{23} }m^{-3}$, and $n$ doping of ${2 \cdot 10^{21} }m^{-3}$.
The other one has a device length of $0.25 \mu m$, an n-channel length of $50nm$,
an $n^+$ doping of ${5 \cdot 10^{24} }m^{-3}$, 
and $n$ doping of ${1 \cdot 10^{21} }m^{-3}$. 
We show simulations for a potential bias of  $ V_{0}=0.5 \, V$.
For the 400 nm channel diode, the number of cells used in the simulations for each of the variables was:
$N_x=120$, $N_r= 80$ and $N_{\mu}=24$. The interval size for $r$ is taken as $\Delta r = 0.45$, having
then $r_{max} = 36$. 
We use a mesh as in \cite{CGMS-CMAME2008} which gives better resolution close to the first juncture at $x = 0.3 \mu m$,
and which also has a finer refinement close to the pole in the direction of the electric field.
It uses $\Delta x = 0.01$ for the first 20 cells in $x$-space, 
$\Delta x = 0.005$ for the next 40 cells, and $\Delta x = 0.01$ for the last 60 cells. Regarding $\mu$,  
it uses 12 cells for $\mu \in [-1,0.7]$, and 12 cells for $\mu \in [0.7, 1]$. 

For the 50 nm channel diode, the number of cells used in the simulations for each of the variables was:
$N_x=64$, $N_r= 80$ and $N_{\mu}=20$. The interval size for $r$ is taken as $\Delta r = 0.8$, having
then $r_{max} = 64$. 
As in \cite{CGMS-CMAME2008}, we use a mesh intended to give better resolution close to the junctures at $x = 0.1 \mu m$ and $x=0.15 \mu m$,
and which also has a finer refinement close to the pole in the direction of the electric field.
It uses $\Delta x = 0.01$ for the first 9 cells in $x$-space, 
$\Delta x = 0.001$ for the next 20 cells close to the the first juncture at $x=0.1\mu m$, 
$\Delta x = 0.005$ for 6 cells at the center of the n-channel, 
$\Delta x = 0.001$ for the next 20 cells close to the second juncture at $x=0.15\mu m$, and $\Delta x = 0.01$ for the last 9 cells. Regarding $\mu$,  it uses 10 cells for $\mu \in [-1,0.7]$, and 10 cells for $\mu \in [0.7, 1]$. 
We let the solver run until $t = 5.0 ps$, a time when the simulations are close to a numerical stationary state. \\

We show plots of the average velocity, the average energy, the momentum (proportional to the current), the 
electric field and potential, for both the $400nm$ channel and $50nm$ channel diodes.
There is a clear quantitative difference, particularly in kinetic moments such as average velocity, average
energy, and momentum (current), whose values depend on the energy band model used in each case then. 
This should be expected since these kinetic moments are averages of  quantities related to $\varepsilon(k)$ or its partial derivatives  in $k$-space.

\begin{figure}
\includegraphics[angle=0,width=0.50\linewidth]{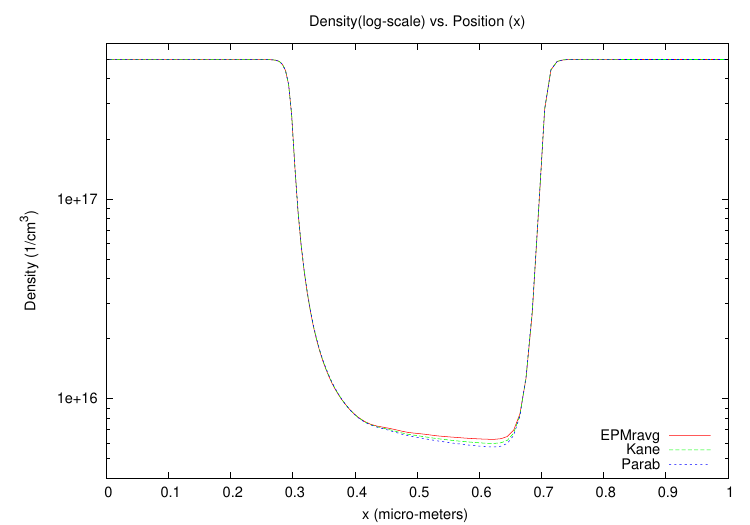}
%
\includegraphics[angle=0,width=0.50\linewidth]{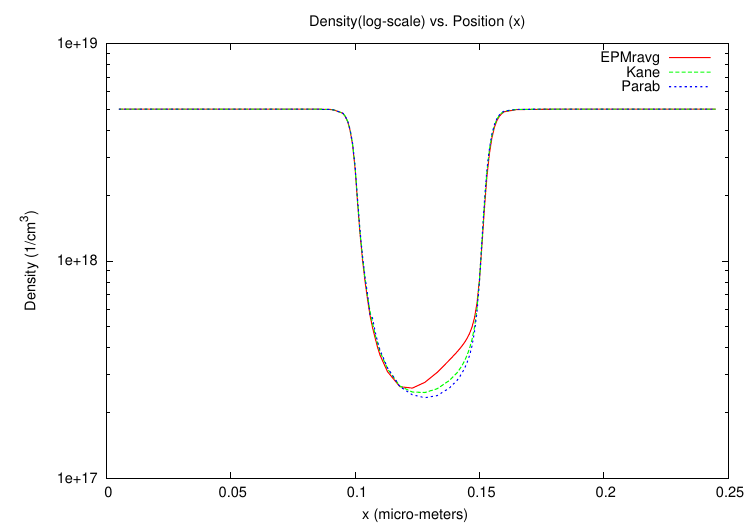}
\caption{Density ($\rho$, in log-scale) vs. position ($x$) plots for different conduction 
band models: parabolic, 
Kane, EPM average. $t=10.0ps$. Left:400nm channel. Right: 50nm channel. 0.5 Volts Bias.}
\end{figure}
\begin{figure}
{\includegraphics[angle=0,width=0.5\linewidth]{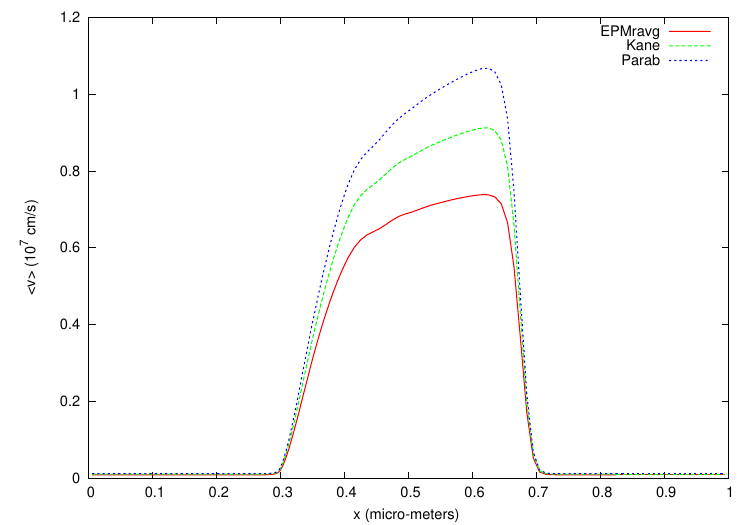}}
%
{\includegraphics[angle=0,width=0.5\linewidth]{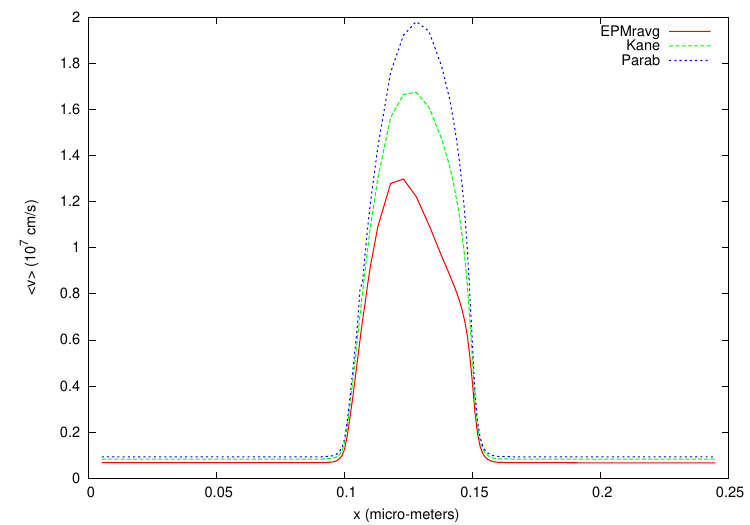}}
\caption{Average velocity ($v$) vs. position ($x$) plots for different 
conduction band models: 
parabolic, Kane, EPM average. $t=10.0ps$. Left:400nm channel. Right: 50nm channel.  0.5 Volts Bias} 
\end{figure}

\begin{figure}
{\includegraphics[angle=0,width=0.50\linewidth]{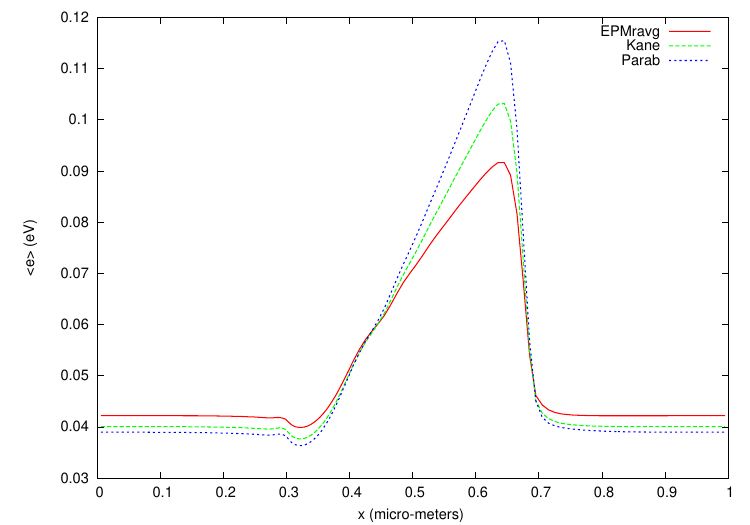}}
%
{\includegraphics[angle=0,width=0.50\linewidth]{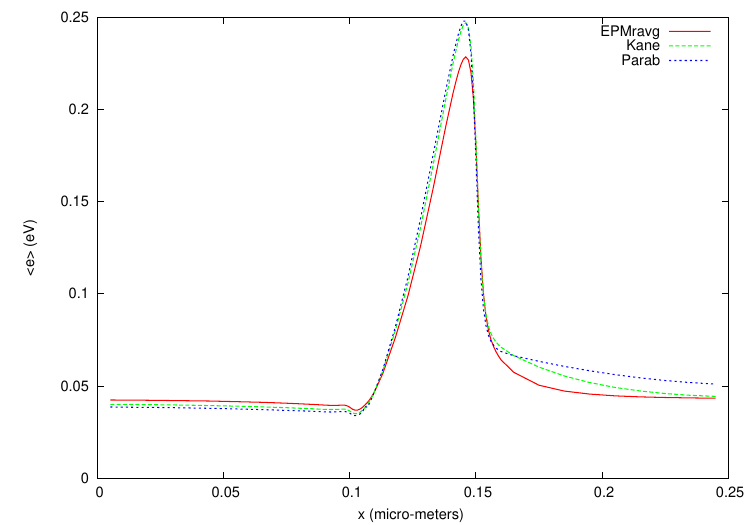}}
\caption{Average energy ($\varepsilon$) vs. position ($x$) for different 
conduction band models: 
parabolic, Kane, EPM average. $t=10.0ps$. Left:400nm channel. Right: 50nm channel. 0.5 Volts Bias.}
\end{figure}
\begin{figure}
{\includegraphics[angle=0,width=0.50\linewidth]{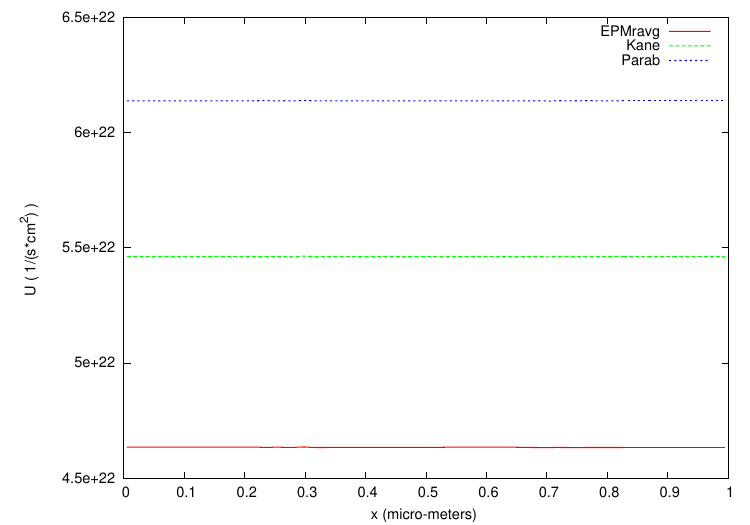}}
%
{\includegraphics[angle=0,width=0.50\linewidth]{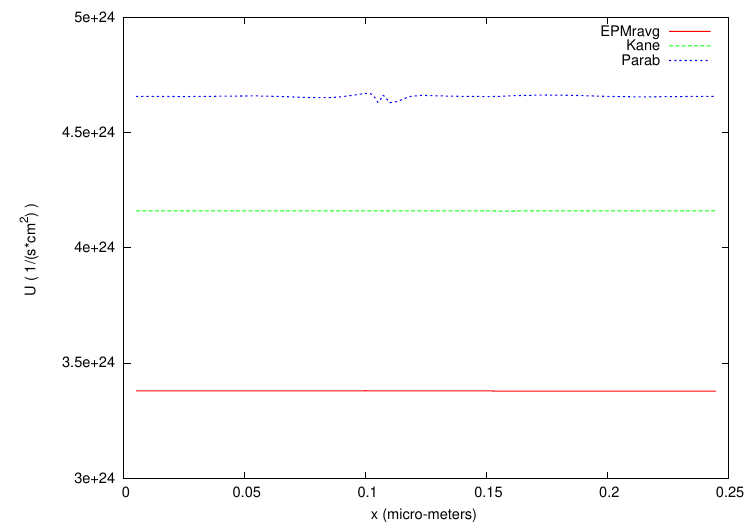}}
\caption{Current (Momentum) vs. position ($x$) for different conduction 
band models: parabolic, 
Kane, EPM average. $t=10.0ps$. Left:400nm channel. Right: 50nm channel. 0.5 Volts Bias.} 
\end{figure}

\begin{figure}
\includegraphics[angle=0,width=0.50\linewidth]{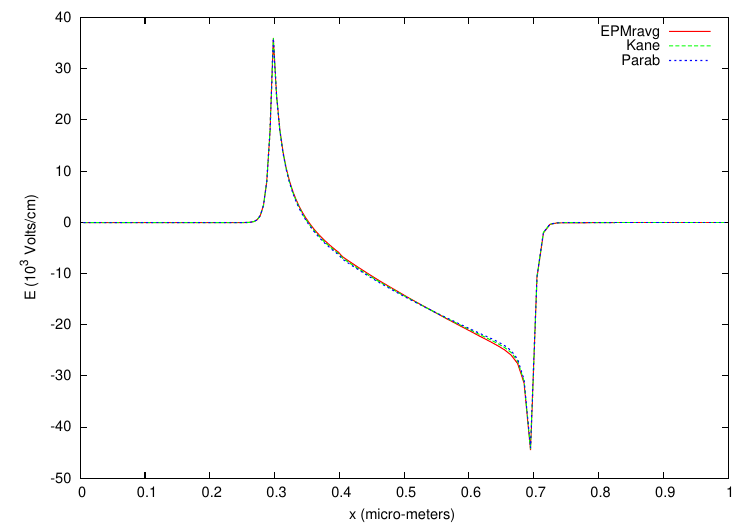}
%
\includegraphics[angle=0,width=0.50\linewidth]{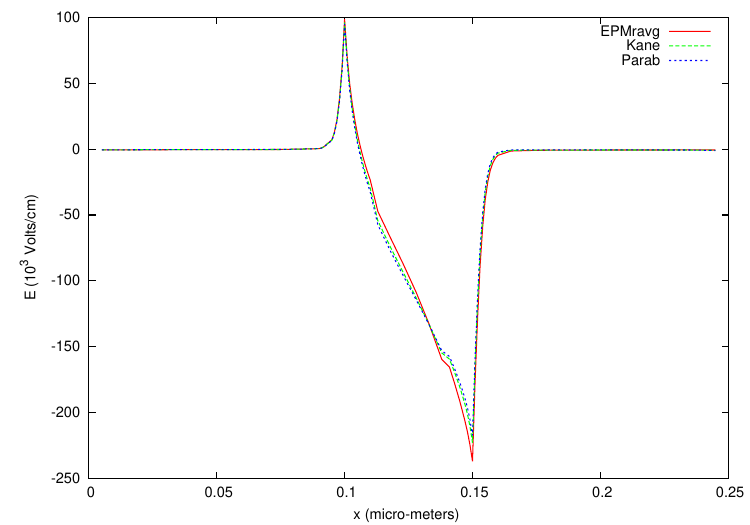}
\caption{Electric field ($E$) vs. position ($x$) plots for different 
conduction band models: 
parabolic, Kane, EPM average. $t=10.0ps$. Left:400nm channel. Right: 50nm channel. 0.5 Volts Bias.}
\end{figure}

\begin{figure}
\includegraphics[angle=0,width=0.50\linewidth]{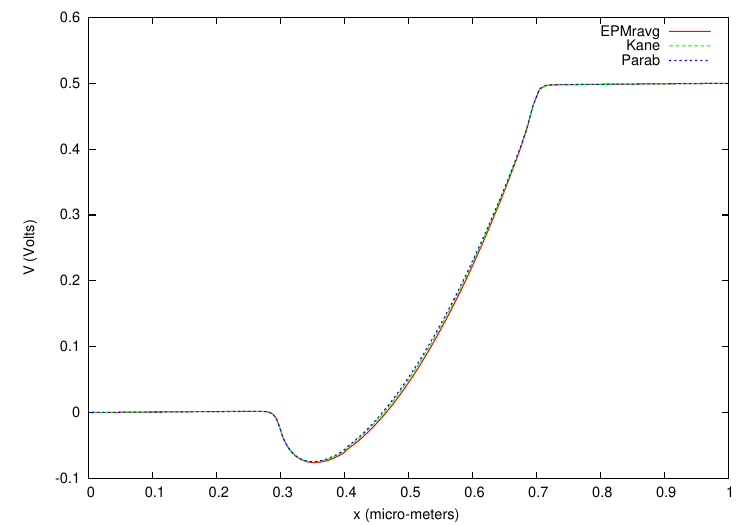}
%
\includegraphics[angle=0,width=0.50\linewidth]{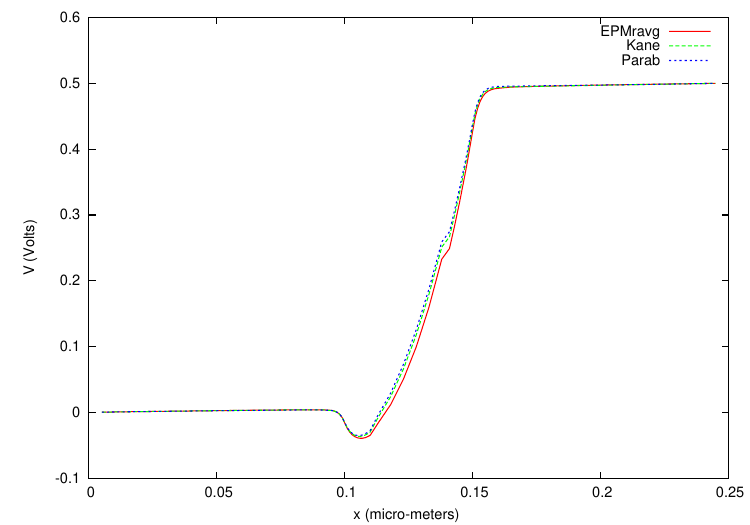}
\caption{Electric Potential ($V$) vs. position ($x$) plots for different 
conduction band models: 
parabolic, Kane, EPM average. $t=10.0ps$. Left:400nm channel. Right: 50nm channel. 0.5 Volts Bias.}
\end{figure}

\begin{figure}
\includegraphics[angle=0,width=1.0\linewidth]{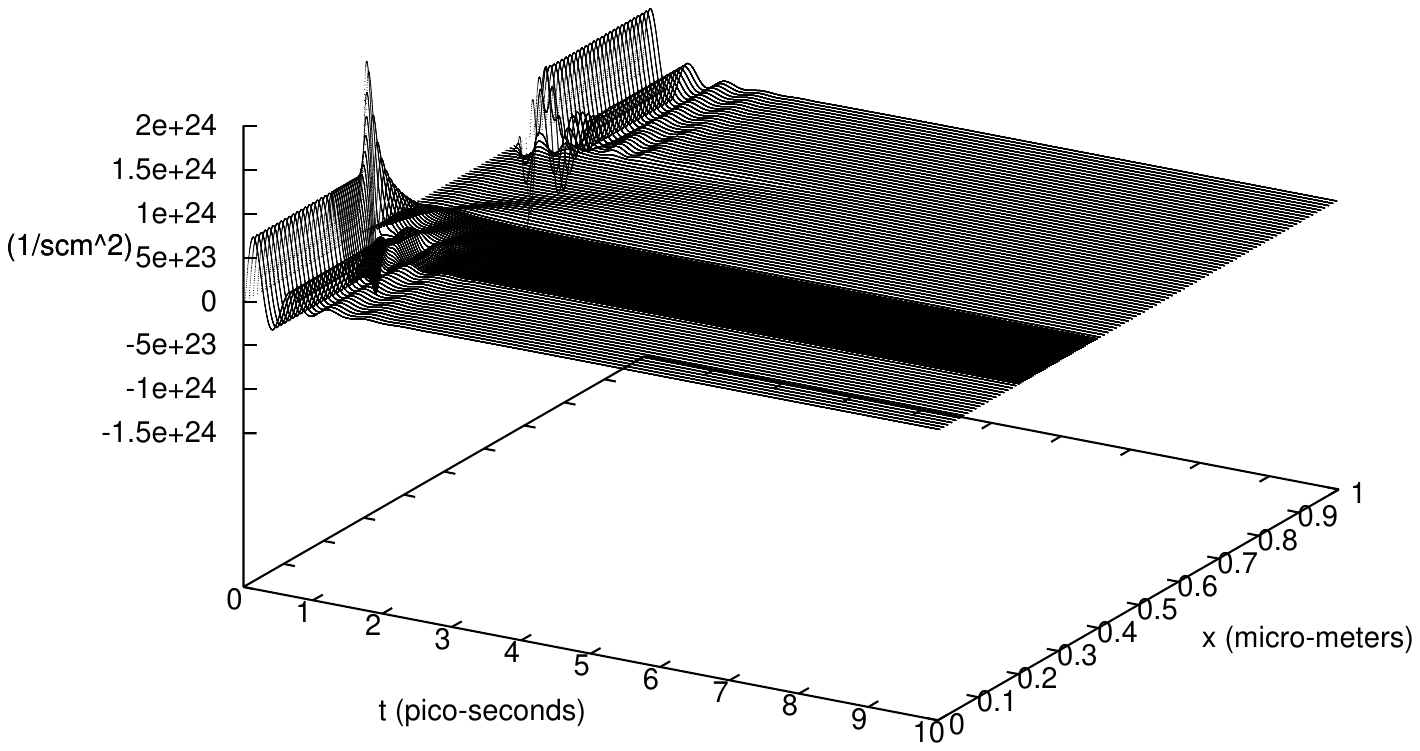} 
\caption{Current (Momentum) vs. $(t,x)$ for the 400nm channel diode, using 
the EPM average band. From initial time to $t=10.0ps$. 0.5 Volts bias.
The initial condition is proportional to the Maxwellian $\exp(-\varepsilon(r))$ times the doping profile $N(x)$.
The initial oscillations are produced by the initial state being far from the final steady state.
} 
\end{figure}

\begin{figure}
\includegraphics[angle=0,width=1.0\linewidth]{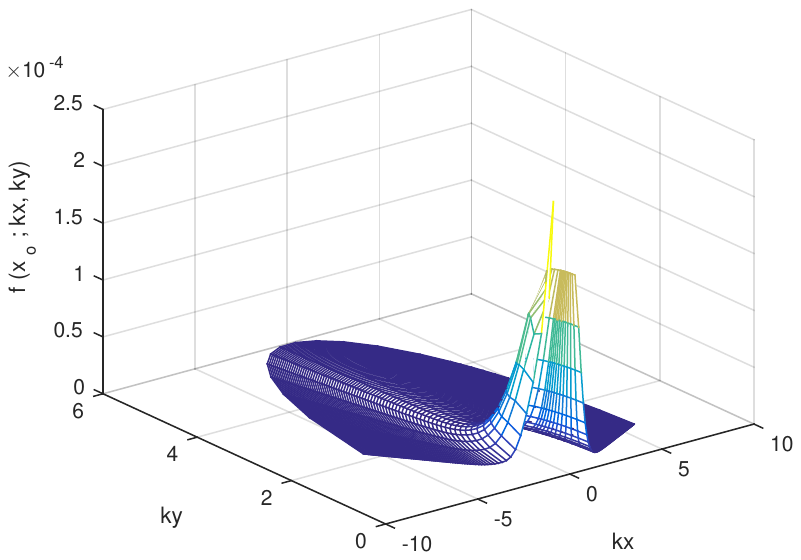} 
\caption{PDF $f(r,\mu;t_0,x_0)$ vs. $(k_x,k_y)$ coordinates (Azimuthal symmetry, with $k_z = 0$)  
at the point 
$x_0=0.3 \mu m$ at $t_0=10.0ps$, for the $1\mu m$ diode with a 400$nm$ channel,
using the EPM radial average energy band. 0.5 Volts bias.} 
\end{figure}

\begin{figure}
\includegraphics[angle=0,width=1.0\linewidth]{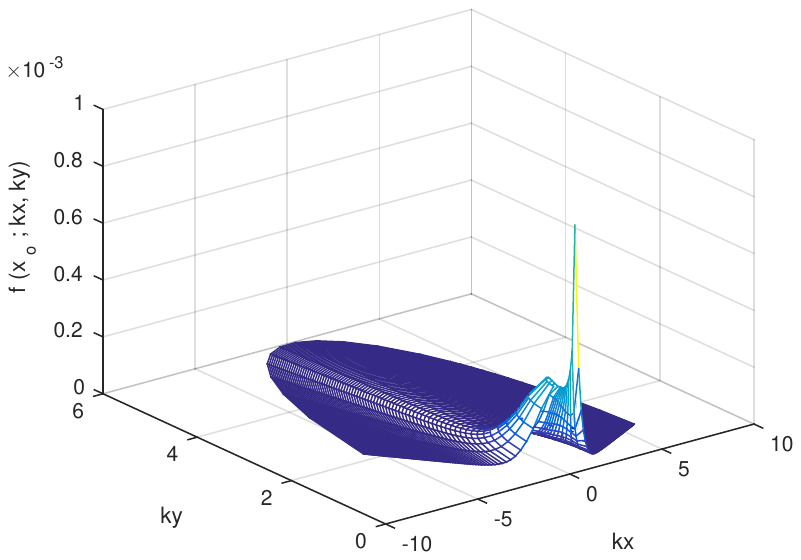} 
\caption{PDF $f(r,\mu;t_0,x_0)$ vs. $(k_x,k_y)$ coordinates (Azimuthal symmetry, with $k_z = 0$) 
at the point 
$x_0=0.7 \mu m$ at $t_0=10.0ps$, for the $1\mu m$ diode with a 400$nm$ channel,
using the EPM radial average energy band. 0.5 Volts bias.} 
\end{figure}

\subsection{Current - Voltage Characteristic Curves}

\begin{figure}
\includegraphics[angle=0,width=0.32\linewidth]{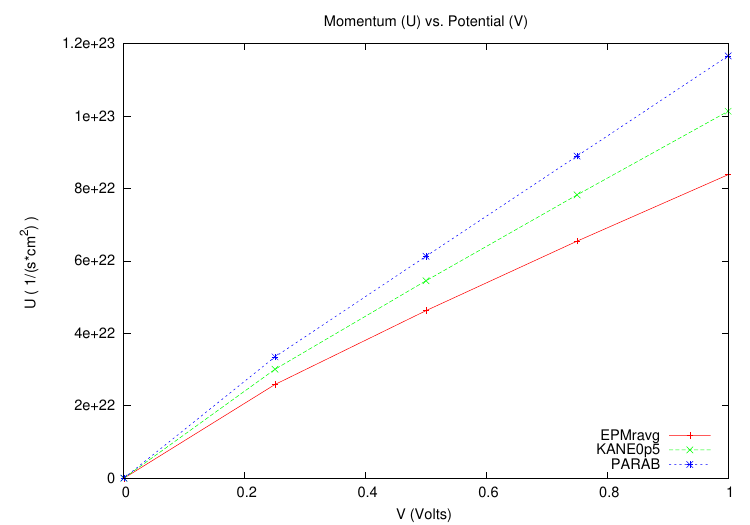}
\includegraphics[angle=0,width=0.32\linewidth]{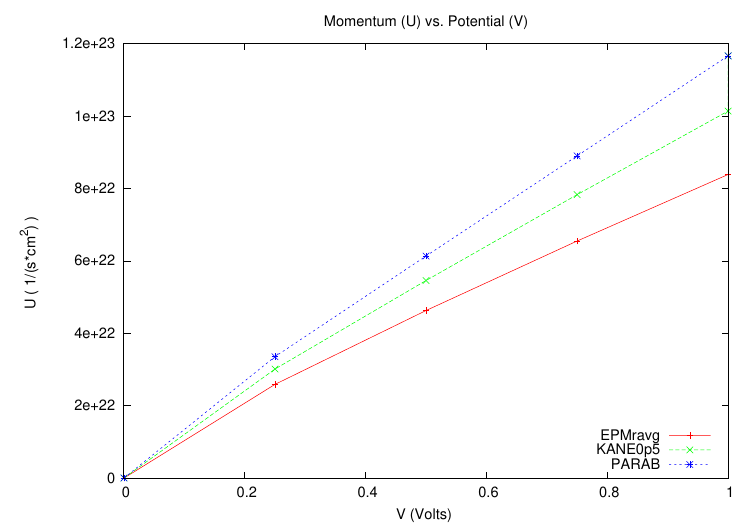}
\includegraphics[angle=0,width=0.32\linewidth]{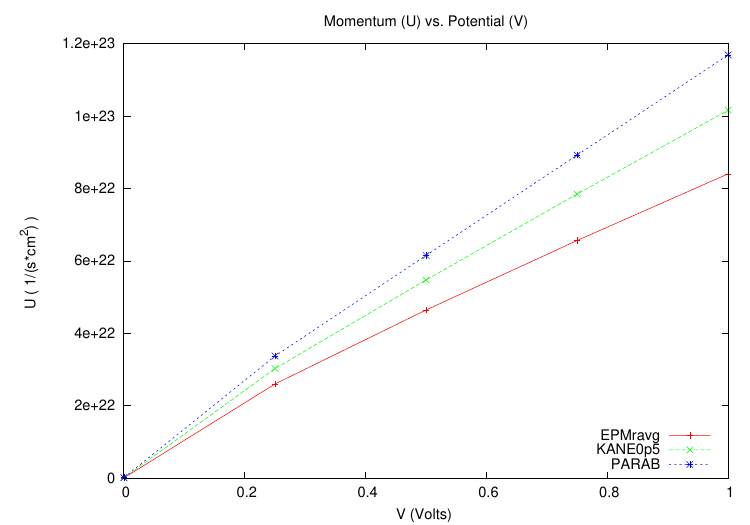}
\caption{ Momentum (U) at $t=10.0ps$ vs. Applied Bias ($V$) plots for voltages of 0., 0.25, 0.5, 0.75, 1.0 V for the 
parabolic, Kane, and EPM average band models. Center: Original Mesh. Left: Refined $r$-Mesh by a factor of 1/2. Right: Coarsened $r$-Mesh by a factor of 2. 400nm channel device. There is a clear quantitative difference trend between the current values for the different bands, with the parabolic band giving current values bigger than the Kane band, and the EPM average band having lower current values than the other models.
 }
\label{fig:IVcurves400}
\end{figure}

\begin{figure}
\includegraphics[angle=0,width=0.32\linewidth]{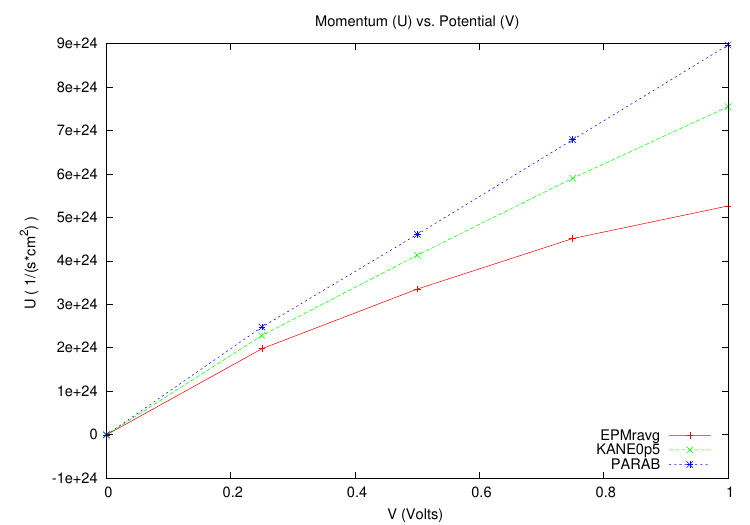}
\includegraphics[angle=0,width=0.32\linewidth]{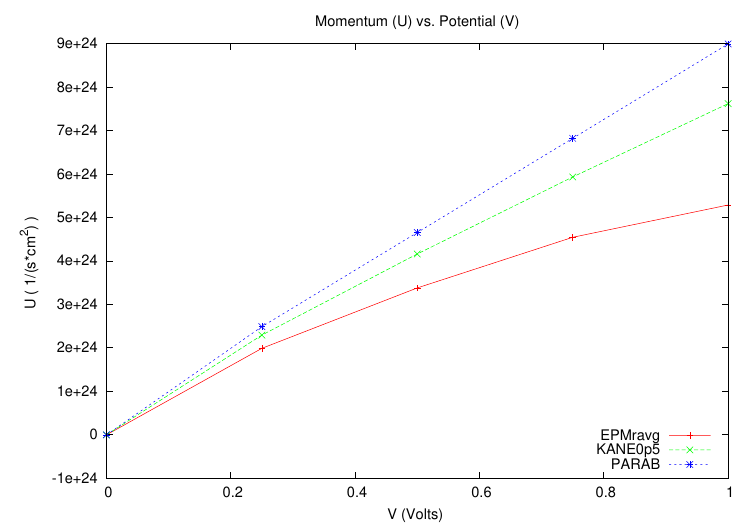}
\includegraphics[angle=0,width=0.32\linewidth]{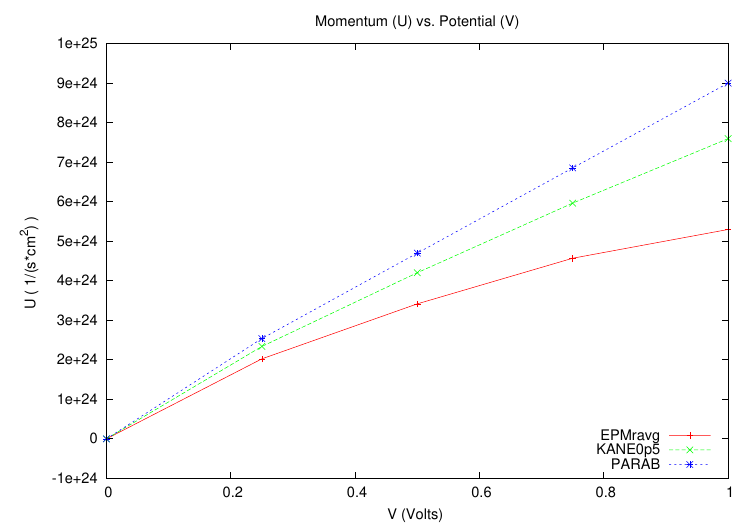}
\caption{ Momentum (U) at $t=10.0ps$ vs. Applied Bias ($V$) plots for voltages of 0., 0.25, 0.5, 0.75, 1.0 V for different 
conduction band models: 
parabolic, Kane, EPM average. Center: Original Mesh. Left: Refined $r$-Mesh by a factor of 1/2. Right: Coarsened $r$-Mesh by a factor of 2. 50nm channel device. There is a clear quantitative difference trend between the current values for the different bands, with the parabolic band giving current values bigger than the Kane band, and the EPM average band having lower current values than the other models.
 }
\label{fig:IVcurves50}
\end{figure}

We also perform a study of the current - voltage characteristic curves for the considered pair of devices with our 3 different band models. 
We show in Fig. \ref{fig:IVcurves400} the plots of the Momentum (U) vs. 
Bias ($V$) for voltages of $0, 0.25, 0.5, 0.75, 1.0$ V with the mesh described in the previous section for the 400nm channel device,
and in Fig. \ref{fig:IVcurves50} the respective plots for the 50nm channel device. We observe a clear quantitative difference between the currents according to the used band model. The currents predicted using the parabolic band are bigger than the ones obtained using the Kane band, and the EPM radial average predicts lower current values than the other two models. We compute as well the IV curves with a refined mesh in the $r$ variable subdividing the intervals of the original mesh in half the size, and with a coarse mesh as well by joining pairs of subsequent intervals, using in all cases the same domain. The current voltage characteristics predicted with the coarser and finer $r$-meshes are quite close to the predictions with the original mesh used. 

\begin{figure}
\includegraphics[angle=0,width=0.95\linewidth]{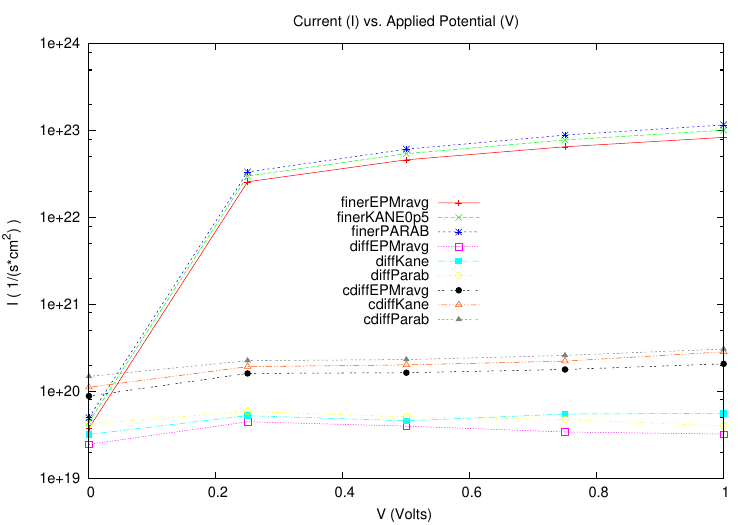}
\caption{ 
Log plot of Momentum (U) at $t=10.0ps$ vs. Applied Bias ($V$) for voltages of 0., 0.25, 0.5, 0.75, 1.0 V for the 
parabolic, Kane, and EPM average band models with a refined $r$-Mesh by a factor of 1/2. Log plots of the difference of these values with the original mesh ($U - U_R$) and with a coarser mesh ($U_C - U_R$). 
400nm channel device. Notice that the order of magnitude of the current values for nonzero voltages is twice bigger than the difference between those currents for the coarser and the finer mesh, and three times bigger than the difference between those currents for the original and final mesh. The differences are consistently lower for the EPM average band
when compared to the Kane and Parabolic bands. 
}
\label{fig:diffIV400}
\end{figure}

\begin{figure}
\includegraphics[angle=0,width=0.95\linewidth]{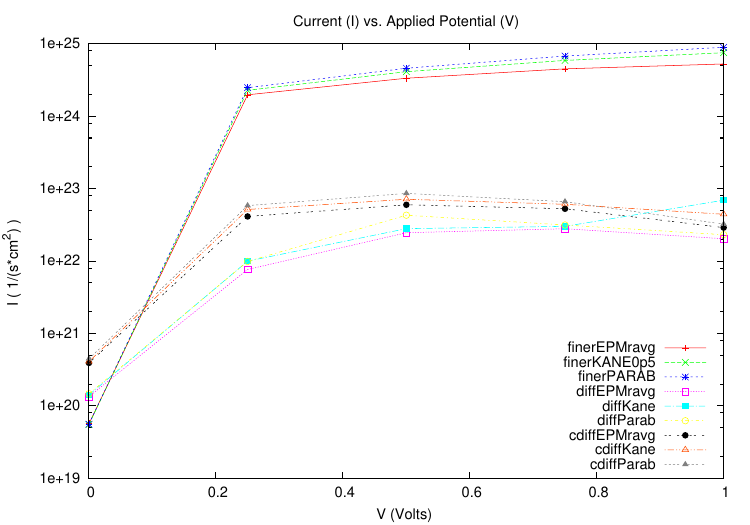}
\caption{ 
Log plot of Momentum (U) at $t=10.0ps$ vs. Applied Bias ($V$) for voltages of 0., 0.25, 0.5, 0.75, 1.0 V for the 
parabolic, Kane, and EPM average band models with a refined $r$-Mesh by a factor of 1/2. Log plots of the difference of these values with the original mesh ($|U - U_R|$) and with a coarser mesh ($|U_C - U_R|$) (these values are only negative for 0 V, therefore the absolute value). 
50nm channel device. Notice that the order of magnitude of the current values for nonzero voltages is twice bigger than the difference between those currents for the coarser and the finer mesh, as for the difference between those currents for the original and final mesh.
The differences are consistently lower for the EPM average band when compared to the Kane and Parabolic Band.
}
\label{fig:diffIV50}
\end{figure}

We compare as well in figures \ref{fig:diffIV400} and \ref{fig:diffIV50}
the momentum predicted with a finer $r$-mesh for our set of bias with the differences of these moments with the predicted ones by our original mesh ($U - U_R$) and a coarser mesh ($U_C - U_R$).
For the 400nm channel, we observe in Fig. \ref{fig:diffIV400} that the order of magnitude of the current values for nonzero voltages is twice bigger than the difference between those currents for the coarser and the finer mesh, and three times bigger than the difference between those currents for the original and final mesh. The differences are consistently lower for the EPM average band when compared to the Kane and Parabolic bands. 

For the 50nm channel, we observe in Fig. \ref{fig:diffIV50}
that the order of magnitude of the current values for nonzero voltages is twice bigger than the difference between those currents for the coarser and the finer mesh, as for the difference between those currents for the original and final mesh. The differences are consistently lower for the EPM average band when compared to the Kane and Parabolic Band.

The set of plots presented for our devices point to several facts.
In addition to the clear quantitative difference between the current voltage characteristics predicted according to which band model is used, with the EPM radial average currents below the Kane predictions and these below the Parabolic currents, the fact that the difference in the IV curves obtained with different meshes is at least two orders of magnitude below the current values indicates that IV curves of comparable quantitative accuracy can be obtained with coarser meshes in the $r$-variable with a reduced computational effort. We also notice that, for nonzero bias, the values of the differences $U-U_F$, $U_C-U_F$ between predicted currents is always positive, that is, the coarsening of the mesh comes with a slight increase of the predicted current. Except for one point in Fig. \ref{fig:diffIV50} related to the Kane band with a $1.0$V bias, the differences $U_C-U_F$ in currents with the coarser mesh are bigger than the difference $U-U_F$ between the current predicted with the original mesh and the refined mesh. Moreover, 
since we observe that both these differences $U_C-U_F$ and $U-U_F$, 
are lower for the EPM average band model, 
the variations in the current prediction are smaller under the coarsening of the mesh when using the EPM average as a band model.

\newpage

\section{Conclusions}
The implementation of a spherical average of an EPM full band structure
as a conduction energy band model in a DG solver for Boltzmann - Poisson
represents a computational strategy that is a midpoint between a radial 
and an anisotropic full band model. 
This difference in the values of the energy band and its derivatives, 
introduced via the spherical average of EPM bandstructure values over $k$-spheres
and the spline interpolation of derivatives 
predicts a quantitative correction in kinetic moments (averages) related to the energy band model, such 
as average velocity, energy, and particularly the momentum (proportional to the current) given by our solver. 
It emphasizes then the importance of an accurate physical modeling of the energy band structure and its partial derivatives
as these functions drive the mechanisms of collision (electron - phonon scattering) and transport (via the electron group velocity)
whose balance is the core modeling of electron transport in semiconductor by the Boltzmann - Poisson system.
This highlights 
the importance of band models and their features (anisotropy, numerical approximation of 
its values and interpolation of their derivatives, for example) 
in the BP numerical modeling of electron transport via DG schemes. 

Future work will focus on the implementation of anisotropic EPM full bands 
in the DG solver for BP, and on developing a positivity-preserving DG numerical scheme, along 
with error estimates for the BP system under consideration.
\section{Appendix}
\appendix

\section{TBE in Divergence Form for $\bk$ in Spherical Coordinates}
The divergence in $\bk$ in the standard spherical coordinates 
$(|\bk|,\theta,\varphi)$ for a vector field $A(\bx,\bk,t) = (A_1,A_2,A_3) = A_{|\bk|} \hat{e}_{|\bk|} 
+ A_{\theta} \hat{e}_{\theta} + A_{\varphi} \hat{e}_{\varphi} $ has the expression

\begin{equation} \label{DivSpherStd}
 \nabla_{\bk} \cdot A = \frac{1}{|\bk|^2} 
 \frac{\partial ( |\bk|^2 A_{|\bk|} )}{\partial |\bk|} + 
 \frac{1}{|\bk| \sin \theta} 
 \frac{\partial \left( A_{\theta} \sin \theta \right) }{\partial \theta} 
 + \frac{1}{|\bk| \sin \theta } \frac{\partial A_{\varphi}}{\partial \varphi}
\end{equation}

The divergence of $A$, with a respective orthogonal decomposition: 
$A = A_{r} \hat{e}_{r} + A_{\mu} \hat{e}_{\mu} + A_{\varphi} \hat{e}_{\varphi} $,
in terms of the modified spherical coordinates $(r,\mu,\varphi)$ used in the TBE
is obtained from (\ref{DivSpherStd}) by taking into account:
$ { |\bk|^2}  = ( {2m^*\kbt}/\hbar^2 ) r $, $\mu = \cos \theta$.  
Therefore 
$\dfrac{dr}{d|\bk|} = \dfrac{ \hbar }{ \sqrt{2m^*\kbt} } \, 2\sqrt{r} \, $, 
$\dfrac{d \mu}{d \theta} = - \sqrt{1 - \mu^2} $,
following that $ \hat{e}_{\mu} = - \hat{e}_{\theta}, \, A_{\mu} = - A_{\theta} $, and $ \hat{e}_{|\bk|} = \hat{e}_{r}, \, A_{|\bk|} = A_{r} \, $.
We have then
\begin{eqnarray*} 
 \nabla_{\bk} \cdot A & = &  
 \frac{1}{r} \frac{\partial ( r A_{r} )}{\partial r} \frac{d r}{d |\bk|} +
 \dfrac{ \hbar }{ \sqrt{2m^*\kbt} }
\left(
 \frac{1}{\sqrt{r} \sin \theta} \frac{\partial ( A_{\theta} \sin \theta )}{\partial \mu} 
 \frac{d \mu}{d \theta} 
 + \frac{1}{\sqrt{r} \sin \theta } \frac{\partial A_{\varphi}}{\partial \varphi}
\right)
\\[7pt]
& = & 
 \dfrac{ \hbar }{ \sqrt{2m^*\kbt} } \cdot
 \frac{2}{\sqrt{r}}  \left[ 
 \frac{\partial }{\partial r} \left( r A_{r} \right) + 
 \frac{ \partial }{\partial \mu} \left( - \frac{ \sqrt{1 - \mu^2}}{2} A_{\theta} \right) 
 + \frac{\partial}{\partial \varphi} \left( \frac{1}{2 \sqrt{1 - \mu^2}} A_{\varphi} \right)
 \right] .
\end{eqnarray*}
We obtain then the divergence in the modified spherical coordinates used in this work: 
\begin{equation} \label{DivSpherRMP} 
 \frac{\sqrt{r}}{2} \,  \nabla_{\bk} \cdot A =  
\left( \dfrac{ \hbar }{ \sqrt{2m^*\kbt} } \right)
\left[
 \frac{\partial}{\partial r}  \left( r A_{r} \right) + 
 \frac{ \partial }{\partial \mu} \left(  \frac{ \sqrt{1 - \mu^2}}{2} A_{\mu} \right)
 + \frac{ \partial}{\partial \varphi} \left( \frac{1}{2 \sqrt{1 - \mu^2}} A_{\varphi} \right)
\right]
\end{equation}
So, the $\bk$-transport term in the TBE (\ref{boltztrans}) can be expressed 
in the divergence form (\ref{DivSpherRMP}) by using:
\begin{equation}
 \frac{\sqrt{r}}{2} \left( \frac{-q \, \bE(t,\bx)}{ \hbar} \cdot \nabla_{\bk} f \right) 
= \frac{\sqrt{r}}{2} \nabla_{\bk} \cdot \left( \frac{-q \, \bE(t,\bx)}{ \hbar} f \right)
= \frac{\sqrt{r}}{2} \nabla_{\bk} \cdot A
\end{equation} 
 for the vector field
\begin{equation} 
  A = \frac{-q \, \bE(t,\bx)}{ \hbar} f  = \frac{-q \, \bE(t,\bx)}{ \hbar} \cdot \frac{\Phi}{\sqrt{r}/2} 
\end{equation}
The formula (\ref{DivSpherRMP})  mentioned above for $A$ can be interpreted geometrically as a flow 
of electric field in the orthogonal directions of the spherical coordinate geometry used, since by 
definition: 
\begin{equation}
A_r = A \cdot \hat{e}_r , \quad A_{\mu} = A \cdot \hat{e}_{\mu}, \quad A_{\varphi} = A \cdot 
\hat{e}_{\varphi} .
\end{equation}

\section{Details of calculation  of the transport terms} 
\label{sec:trans}

The following notation for boundary terms will be needed. We denote by 

\begin{eqnarray}
 \hat{\Phi}_{i \pm \frac{1}{2}} & = & \hat{\Phi}(t,x_{i \pm \frac{1}{2}}, y, r, \mu, \varphi) \, , \quad
 \hat{\Phi}_{j \pm \frac{1}{2}} = \hat{\Phi}(t,x, y_{j \pm \frac{1}{2}}, r, \mu, \varphi) \\
 \hat{\Phi}_{k \pm \frac{1}{2}} & = & \hat{\Phi}(t,x, y, r_{k \pm \frac{1}{2}}, \mu, \varphi) \, , \quad
 \hat{\Phi}_{m \pm \frac{1}{2}} = \hat{\Phi}(t,x, y, r, \mu_{m \pm \frac{1}{2}}, \varphi) \, , \quad 
 \hat{\Phi}_{n \pm \frac{1}{2}} = \hat{\Phi}(t,x, y, r, \mu, \varphi_{n \pm \frac{1}{2}}) \nonumber\\
\eta^{p}_{i \pm 1,j}|_{i \pm \ot} & = & \eta^{p}_{i \pm 1,j}(x_{i \pm \ot},y) \, , \quad \eta^{p}_{i,j}|_{i \pm \ot} = \eta^{p}_{i,j}(x_{i \pm \ot},y) \, , \quad
\\
\eta^{p}_{i,j \pm 1}|_{j \pm \ot} & = & \eta^{p}_{i ,j \pm 1}(x,y_{j \pm \ot}) \, , \quad \eta^{p}_{i,j}|_{j \pm \ot} = \eta^{p}_{i,j}(x,y_{j \pm \ot}) \, , \quad
p \in \{ 0, \, 1, \, \cdots \, , 5 \} . \nonumber\\
\xi^{p}_{k,m,n}|_{k \pm \ot} & = & \xi^{p}_{k,m,n}(r_{k \pm \ot},\mu,\varphi) \, , \quad 
\xi^{p}_{k,m,n}|_{m \pm \ot}  =  \xi^{p}_{k,m,n}(r, \mu_{m \pm \ot},\varphi) \, , \quad 
\xi^{p}_{k,m,n}|_{n \pm \ot}  =  \xi^{p}_{k,m,n}(r,\mu,\varphi_{n \pm \ot}) \quad \nonumber\\
\xi^{p}_{k \pm 1,m,n}|_{k \pm \ot} & = & \xi^{p}_{k \pm 1,m,n}(r_{k \pm \ot},\mu,\varphi) \, , \quad 
\xi^{p}_{k,m\pm 1,n}|_{m \pm \ot}  =  \xi^{p}_{k,m\pm 1,n}(r, \mu_{m \pm \ot},\varphi) \, , \\
\xi^{p}_{k,m,n\pm 1}|_{n \pm \ot}  & = &  \xi^{p}_{k,m,n\pm 1}(r,\mu,\varphi_{n \pm \ot}) \, , \quad p \in \{ 0, \, 1, \, \cdots \, , 5 \}. \nonumber
\end{eqnarray}

We consider first the weak formulation of the transport terms in space, namely $ \frac{\partial}{\partial x} \left( a_1 \Phi \right) $ and $ \frac{\partial}{\partial y} \left( a_2 \Phi \right) $, related to the advection in $\bx$, where the first cartesian component $a_1$ of the electron group velocity is involved. Because their discretization forms are similar we only present the one for $ \frac{\partial}{\partial x} \left( a_1 \Phi \right) $.

{\small
\begin{eqnarray*}
\hspace{-20pt} 
(A_{1}) 
& = &
\int_{\Omega_{I}}
\frac{\partial \mbox{ }}{\partial x} 
\left[ a_{1}(r,\mu,\varphi) \, \Phi(t,\vec{x},\vec{r}) \right] 
\psi(\vec{x},\vec{r}) \: d\vec{x} \, d\vec{r} \, = \\
& = & 
\dfrac{1}{\Delta x_{i}}
\int_{\Omega_{I}}  
a_{1}(\vec{r})
\left[ 
\hat{\Phi}_{i + \ot} 
\psi_{i + \ot} 
-
\hat{\Phi}_{i - \ot} 
\psi_{i - \ot} \right]  
d\vec{x} d\vec{r}
-
\int_{\Omega_{I}}
a_{1}(\vec{r}) \Phi(t,\vec{x},\vec{r})
\frac{\partial \mbox{ }}{\partial x} 
\psi(\vec{x},\vec{r}) d\vec{x} d\vec{r} \, .
\end{eqnarray*}
}

Due to the upwind flux rule, we have to consider two cases 
depending on the sign of $a_1$. In the sequel, the symbol $\approx$ will denote the approximation of given integral terms

If $a_{1}(\vec{r}) > 0 $ in $K_{kmn}$, for $q = 0,...,5$, one obtains
{\small
\begin{eqnarray*}
(A_{1}) & \approx &
\sum_{p=0}^{5} 
\dfrac{W_{I}^p (t)}{\Delta x_{i}}
\int_{\Omega_{I}} 
a_{1}(\vec{r})  \eta^{p}_{i,j}|_{i + \ot}  \xi^{p}_{k,m,n}(\vec{r}) 
\eta^{q}_{i,j}|_{i + \ot} \xi^{q}_{k,m,n}(\vec{r}) 
d\vec{x} d\vec{r}
\\
& - &
\sum_{p=0}^{5} 
\dfrac{W_{i-1 \, jkmn}^p (t)}{\Delta x_{i}}
\int_{\Omega_{I}} 
a_{1}(\vec{r})  \eta^{p}_{i-1,j}|_{i - \ot}  \xi^{p}_{k,m,n}(\vec{r}) 
\eta^{q}_{i,j}|_{i - \ot} \xi^{q}_{k,m,n}(\vec{r}) 
d\vec{x} d\vec{r}
\\
& - &
\sum_{p=0}^{5} W_{I}^p (t)
\int_{\Omega_{I}}
a_{1}(\vec{r}) \, \eta^{p}_{i,j}(\vec{x}) \, \xi^{p}_{k,m,n}(\vec{r}) \,
\frac{2 \, \delta_{q 4}}{\Delta x_{i}} \, \xi^{q}_{k,m,n}(\vec{r}) \: 
d\vec{x} \, d\vec{r}  
\\
& = &
\sum_{p=0}^{5} 
\int_{y_{j - \ot}}^{y_{j + \ot}}
  \eta^{p}_{i,j}|_{i + \ot} \eta^{q}_{i,j}|_{i + \ot}  dy
\cdot
\int_{K_{kmn}}
a_{1}(\vec{r}) \xi^{p}_{k,m,n}(\vec{r}) 
 \xi^{q}_{k,m,n}(\vec{r}) d\vec{r}
\cdot
W_{I}^p (t)
\\
& - &
\sum_{p=0}^{5} 
\int_{y_{j - \ot}}^{y_{j + \ot}}
 \eta^{p}_{i-1,j}|_{i - \ot} \eta^{q}_{i,j}|_{i - \ot} dy
\int_{K_{kmn}}
a_{1}(\vec{r}) \xi^{p}_{k,m,n}(\vec{r}) 
 \xi^{q}_{k,m,n}(\vec{r}) d\vec{r}
\cdot
W_{i-1 \, jkmn}^p(t)
\\
& - &
\sum_{p=0}^{5} 
\left[ 
\frac{2 \, \delta_{q 4}}{\Delta x_{i}}
\int_{x_{i - \ot}}^{x_{i + \ot}}
\int_{y_{j - \ot}}^{y_{j + \ot}}
 \eta^{p}_{i,j}(\vec{x}) \:d\vec{x}
\right] 
\left[ 
\int_{K_{kmn}}
a_{1}(\vec{r}) \xi^{p}_{k,m,n}(\vec{r}) 
 \xi^{q}_{k,m,n}(\vec{r}) \: d\vec{r}
\right] 
\cdot
W_{I}^p (t) \, .
\end{eqnarray*}
}
If $a_{1}(\vec{r}) < 0 $ in $ K_{kmn}$, for $q = 0,...,5$, 
{\small
\begin{eqnarray*}
(A_{1})
&
\approx
&
\sum_{p=0}^{5} 
\dfrac{W_{i+1 \, jkmn}^p(t)}{\Delta x_{i}}
\int_{\Omega_{I}} 
a_{1}(\vec{r}) \eta^{p}_{i+1,j}|_{i + \ot} \xi^{p}_{k,m,n}(\vec{r})
\eta^{q}_{i,j}|_{i + \ot} \xi^{q}_{k,m,n}(\vec{r}) \:
d\vec{x} d\vec{r}
\\
& - &
\dfrac{1}{\Delta x_{i}}
\sum_{p=0}^{5} W_{I}^p (t)
\int_{\Omega_{I}} 
a_{1}(\vec{r})  \eta^{p}_{i,j}|_{i - \ot}  \xi^{p}_{k,m,n}(\vec{r}) 
\eta^{q}_{i,j}|_{i - \ot} \xi^{q}_{k,m,n}(\vec{r}) \:
d\vec{x} d\vec{r}
\\
& -  &
\sum_{p=0}^{5} W_{I}^p (t)
\int_{\Omega_{I}}
a_{1}(\vec{r}) \, \eta^{p}_{i,j}(\vec{x}) \, \xi^{p}_{k,m,n}(\vec{r}) \,
\frac{2 \, \delta_{q 4}}{\Delta x_{i}} \, \xi^{q}_{k,m,n}(\vec{r}) \: 
d\vec{x} \, d\vec{r}  
\\
& = &
\sum_{p=0}^{5} 
\int_{y_{j - \ot}}^{y_{j + \ot}}
 \eta^{p}_{i+1,j}|_{i + \ot}  \eta^{q}_{i,j}|_{i + \ot}  dy
\int_{K_{kmn}}
a_{1}(\vec{r})  \xi^{p}_{k,m,n}(\vec{r}) 
 \xi^{q}_{k,m,n}(\vec{r}) d\vec{r}
\:
W_{i+1 \, jkmn}^p(t)
\\
& -  &
\sum_{p=0}^{5} 
\int_{y_{j - \ot}}^{y_{j + \ot}}
 \eta^{p}_{i,j}|_{i - \ot}  \eta^{q}_{i,j}|_{i - \ot}  dy
\int_{K_{kmn}}
a_{1}(\vec{r})  \xi^{p}_{k,m,n}(\vec{r}) 
 \xi^{q}_{k,m,n}(\vec{r})  d\vec{r}
\:
W_{I}^p (t)
\\
& - &
\sum_{p=0}^{5} 
\frac{2 \, \delta_{q 4}}{\Delta x_{i}} \,
\int_{x_{i - \ot}}^{x_{i + \ot}}
\int_{y_{j - \ot}}^{y_{j + \ot}}
 \eta^{p}_{i,j}(\vec{x}) \:dx \, dy
\int_{K_{kmn}}
a_{1}(\vec{r}) \, \xi^{p}_{k,m,n}(\vec{r}) \,
 \xi^{q}_{k,m,n}(\vec{r}) \: d\vec{r} 
\:
W_{I}^p (t) \, .
\end{eqnarray*}
}

We consider now the weak formulation for the transport terms in momentum space 
$
\frac{\partial \mbox{}}{\partial r} \left( a_{4} \, \Phi \right)  + \frac{\partial \mbox{ }}{\partial \mu} \left( a_{5} \,
\Phi \right) + \frac{\partial \mbox{ }}{\partial \varphi} \left(
a_{6} \, \Phi \right) 
$
advected by the electric field. It can be noticed in Eq.
\ref{boltztrans}
that all the terms $a_4, a_5, a_6$ are the sum of terms of the form 
$a_{*}(\vec{r}) \, E_{*}(t,\vec{x})$,
where
$E_{*}(t,\vec{x})$ is a cartesian component of the electric field. 

The $r$-derivative including $a_{4}$ can be split as a sum of terms as the following 
\begin{eqnarray*}
 (A_{4*}) & = & 
\int_{\Omega_{I}}
\frac{\partial }{\partial r} 
\left[ a_{*}(\vec{r}) \, E_{*}(t,\vec{x}) \, \Phi(t,\vec{x},\vec{r}) \right] 
\psi(\vec{x},\vec{r}) \: d\vec{x} \, d\vec{r}  
\\
& = & 
\dfrac{1}{\Delta r_{k}}
\int_{\Omega_{I}} 
a_{*}|_{k + \ot} \, E_{*}(t,\vec{x}) \, \hat{\Phi}_{k + \ot} \,
\psi_{k + \ot} \:  d\vec{x} \, d\vec{r}  
-
\dfrac{1}{\Delta r_{k}}
\int_{\Omega_{I}}
a_{*}|_{k - \ot} \, E_{*}(t,\vec{x}) \, \hat{\Phi}_{k - \ot} \,
\psi_{k - \ot} \:  d\vec{x} \, d\vec{r} 
\\
& - & 
\int_{\Omega_{I}}
a_{*}(\vec{r}) \, E_{*}(t,\vec{x}) \, \Phi(t,\vec{x},\vec{r})
\frac{\partial \mbox{ }}{\partial r} 
\psi(\vec{x},\vec{r}) \: d\vec{x} \, d\vec{r}  \, .
\end{eqnarray*}
By the Upwind Flux Rule, if $a_{*}(\vec{r}) \, E_{*}(t,\vec{x}) > 0$ in $\partial_r^{\pm} \Omega_{I}$, 
for $q = 0, ...,5 \, ,$ 
{\small
\begin{eqnarray*}
 \hspace{-10pt} (A_{4*}) \hspace{-5pt} 
& \approx &
\sum_{p=0}^{5}
\dfrac{W^p_{I}}{\Delta r_{k}}
\int_{\Omega_{I}} 
a_{*}|_{k + \ot}  E_{*}(t,\vec{x})  \eta^{p}_{i,j}(\vec{x})  \xi^{p}_{k,m,n}|_{k + \ot} 
\eta^{q}_{i,j}(\vec{x})  \xi^{q}_{k,m,n}|_{k + \ot} 
d\vec{x} d\vec{r}
\\
& - &
\sum_{p=0}^{5} \hspace{-4pt}
\dfrac{W^p_{ijk-1  mn}}{\Delta r_{k}}
\int_{\Omega_{I}} 
a_{*}|_{k - \ot} E_{*} \eta^{p}_{i,j}(\vec{x})  
\xi^{p}_{k-1,m,n}|_{k - \ot}
\eta^{q}_{i,j}(\vec{x})  \xi^{q}_{k,m,n}|_{k - \ot} 
d\vec{x}  d\vec{r}
\\
&- &
\sum_{p=0}^{5} W_{I}^p (t)
\int_{\Omega_{I}}
a_{*}(\vec{r}) \, E_{*}(t,\vec{x}) \, \eta^{p}_{i,j}(\vec{x}) \, \xi^{p}_{k,m,n}(\vec{r})
\, \dfrac{2 \, \delta_{q1}}{\Delta r_{k}} \, \eta^{q}_{i,j}(\vec{x}) \:
d\vec{x} \, d\vec{r} 
\\
& = &  
\sum_{p=0}^{5} \hspace{-4pt} 
\left[ 
\int_{x_{i - \ot}}^{x_{i + \ot}} \hspace{-5pt}
\int_{y_{j - \ot}}^{y_{j + \ot}} \hspace{-5pt}
E_{*}(t,\vec{x})  \eta^{p}_{i,j}(\vec{x})  \eta^{q}_{i,j}(\vec{x}) dx  dy 
\right]  
\hspace{-4pt}
\left\lbrace 
\left[ 
\int_{\mu_{m - \ot}}^{\mu_{m + \ot}} \hspace{-5pt}
\int_{\varphi_{n - \ot}}^{\varphi_{n + \ot}} \hspace{-5pt}
a_{*}|_{k + \ot}  \xi^{p}_{k,m,n}|_{k + \ot} 
\xi^{q}_{k,m,n}|_{k + \ot} d\mu d\varphi 
\right.
\right.
\\
& - & \left. 
 \dfrac{2 \, \delta_{q1}}{\Delta r_{k}} \,  \int_{K_{kmn}}
a_{*}(\vec{r}) \,  \xi^{p}_{k,m,n}(\vec{r}) \: d\vec{r}  
\right] 
W_{I}^p 
\left. 
- \left[ 
\int_{\mu_{m - \ot}}^{\mu_{m + \ot}} 
\int_{\varphi_{n - \ot}}^{\varphi_{n + \ot}} 
a_{*}|_{k - \ot} \, \xi^{p}_{k-1,m,n}|_{k - \ot}\, 
\xi^{q}_{k,m,n}|_{k - \ot} \: d\mu \, d\varphi 
\right] 
 W_{ijk-1  mn}^p
\right\rbrace \, .
\end{eqnarray*}
}

\ \\
If $a_{*}(\vec{r}) \, E_{*}(t,\vec{x}) < 0$ in $\partial^{\pm}_r \Omega_{I}$,  
 $q = 0, ...,5 \, ,$
{\small
\begin{eqnarray*}
(A_{4*})
& \approx &
\dfrac{1}{\Delta r_{k}}
\sum_{p=0}^{5} W_{ijk+1 mn}^p(t)
\int_{\Omega_{I}}
a_{*}|_{k + \ot} \, E_{*}(t,\vec{x}) \, \eta^{p}_{i,j}(\vec{x}) \,
 \xi^{p}_{k+1,m,n}|_{k + \ot}
\eta^{q}_{i,j}(\vec{x}) \, \xi^{q}_{k,m,n}|_{k + \ot} \:
d\vec{x} \, d\vec{r} 
\\
& - &
\mbox{ } 
\sum_{p=0}^{5}  
\dfrac{W^p_{I}}{\Delta r_{k}}
\int_{\Omega_{I}} 
a_{*}|_{k - \ot} E_{*}
\eta^{p}_{i,j}(\vec{x})  
\xi^{p}_{k,m,n}|_{k - \ot} 
\eta^{q}_{i,j}(\vec{x}) \xi^{q}_{k,m,n}|_{k - \ot} 
d\vec{x} d\vec{r}
\\
& - &
\mbox{ } 
\sum_{p=0}^{5} W_{I}^p (t)
\int_{\Omega_{I}}
a_{*}(\vec{r}) \, E_{*}(t,\vec{x}) \, \eta^{p}_{i,j}(\vec{x}) \, \xi^{p}_{k,m,n}(\vec{r})
\, \dfrac{2 \, \delta_{q1}}{\Delta r_{k}} \, \eta^{q}_{i,j}(\vec{x}) \:
d\vec{x} \, d\vec{r} 
\\
& = &  \sum_{p=0}^{5} 
\left[ 
\int_{x_{i - \ot}}^{x_{i + \ot}} 
\int_{y_{j - \ot}}^{y_{j + \ot}} 
E_{*}(t,\vec{x}) \, \eta^{p}_{i,j}(\vec{x}) \, \eta^{q}_{i,j}(\vec{x}) \:dx \, dy
\right]  
\\
& & 
\times
\left\lbrace
\left[ 
\int_{\mu_{m - \ot}}^{\mu_{m + \ot}}  
\int_{\varphi_{n - \ot}}^{\varphi_{n + \ot}} 
a_{*}|_{k + \ot} \, \xi^{p}_{k+1,m,n}|_{k + \ot} \, 
\xi^{q}_{k,m,n}|_{k + \ot} \: d\mu \, d\varphi 
\right]
W_{ijk+1 mn}^p(t)
\right.
\\
& & \mbox{} 
- \left[ 
\int_{\mu_{m - \ot}}^{\mu_{m + \ot}} 
\int_{\varphi_{n - \ot}}^{\varphi_{n + \ot}} 
a_{*}|_{k - \ot} \, \xi^{p}_{k,m,n}|_{k - \ot} \, 
\xi^{q}_{k,m,n}|_{k - \ot} \: d\mu \, d\varphi 
\right. 
\left. \left. \mbox{} 
+
\dfrac{2 \, \delta_{q1}}{\Delta r_{k}} \,  \int_{K_{kmn}}
a_{*}(\vec{r}) \,  \xi^{p}_{k,m,n}(\vec{r}) \: d\vec{r}  
\right] 
W_{I}^p (t)
\right\rbrace \, .
\end{eqnarray*}
}

The weak form for the term related to $\frac{\partial}{\partial \mu} \left( a_5 \Phi \right) $ is  

$$
(A_{5*}) =
\int_{\Omega_{I}}
\frac{\partial \mbox{ }}{\partial \mu} 
\left[ a_{*}(\vec{r}) \, E_{*}(t,\vec{x}) \, \Phi(t,\vec{x},\vec{r}) \right] 
\psi(\vec{x},\vec{r}) \: d\vec{x} \, d\vec{r}  \, .
$$

By the Upwind Flux rule, if $a_{*}(\vec{r}) \, E_{*}(t,\vec{x}) > 0$ in $\partial_{\mu}^{\pm} \Omega_{I}$  
, $q = 0, ...,5 \, ,$

{\small
\begin{eqnarray*}
(A_{5*}) \hspace{-5pt} 
&
\approx
&
\sum_{p=0}^{5} 
\left[ 
\int_{x_{i - \ot}}^{x_{i + \ot}} 
\int_{y_{j - \ot}}^{y_{j + \ot}} 
E_{*}(t,\vec{x}) \, \eta^{p}_{i,j}(\vec{x}) \, \eta^{q}_{i,j}(\vec{x}) \:dx \, dy 
\right] 
\\
& &
\mbox{} 
\times \left\lbrace 
\left[ 
\int_{r_{k - \ot}}^{r_{k + \ot}}
\int_{\varphi_{n - \ot}}^{\varphi_{n + \ot}}
a_{*}|_{m + \ot} \, \xi^{p}_{k,m,n}|_{m + \ot} \, 
\xi^{q}_{k,m,n}|_{m + \ot} \: dr \, d\varphi 
\right.
\right.
\left.
- \dfrac{2 \, \delta_{q2}}{\Delta \mu_{m}} \,  \int_{K_{kmn}}
a_{*}(\vec{r}) \,  \xi^{p}_{k,m,n}(\vec{r}) \: 
d\vec{r}  
\right] 
 W_{I}^p (t)
\\
&&
\left.
- \left[ 
\int_{r_{k - \ot}}^{r_{k + \ot}} 
\int_{\varphi_{n - \ot}}^{\varphi_{n + \ot}} 
a_{*}|_{m - \ot} \, \xi^{p}_{k,m-1,n}|_{m - \ot} \, 
\xi^{q}_{k,m,n}|_{m - \ot} \: dr \, d\varphi 
\right] 
 W_{ijkm-1  n}^p(t) \right\rbrace \, .
\end{eqnarray*}
}%

If $a_{*}(\vec{r}) \, E_{*}(t,\vec{x}) < 0$ in $\partial_{\mu}^{\pm}  \Omega_{I}$ , 
$q = 0, ...,5 \, ,$
{\small
\begin{eqnarray*}
(A_{5*})
&
\approx
&
\sum_{p=0}^{5} 
\left[ 
\int_{x_{i - \ot}}^{x_{i + \ot}} 
\int_{y_{j - \ot}}^{y_{j + \ot}} 
E_{*}(t,\vec{x}) \, \eta^{p}_{i,j}(\vec{x}) \, \eta^{q}_{i,j}(\vec{x}) \:dx \, dy 
\right] 
\\
& & 
\times
\left\lbrace 
\left[ 
\int_{r_{k - \ot}}^{r_{k + \ot}}
\int_{\varphi_{n - \ot}}^{\varphi_{n + \ot}} 
a_{*}|_{m + \ot} \, \xi^{p}_{k,m+1,n}|_{m + \ot} \, 
\xi^{q}_{k,m,n}|_{m + \ot} \: dr \, d\varphi 
\right]
W_{ijkm+1 n}^p(t)
\right.
\\
& & 
- \left[ 
\int_{r_{k - \ot}}^{r_{k + \ot}} 
\int_{\varphi_{n - \ot}}^{\varphi_{n + \ot}} 
a_{*}|_{m - \ot} \, \xi^{p}_{k,m,n}|_{m - \ot} \, 
\xi^{q}_{k,m,n}|_{m - \ot} \: dr \, d\varphi 
\right. 
\left. \left. 
+ \dfrac{2 \, \delta_{q2}}{\Delta \mu_{m}} \,  \int_{K_{kmn}}
a_{*}(\vec{r}) \,  \xi^{p}_{k,m,n}(\vec{r}) \: d\vec{r}  
\right] 
W_{I}^p (t)
\right\rbrace \, .
\end{eqnarray*}
}%

Weak form for the term related to $\frac{\partial}{\partial \varphi} \left( a_6 \Phi \right) $ is

\begin{equation}
(A_{6*}) \quad
\int_{\Omega_{I}}
\frac{\partial \mbox{ }}{\partial \varphi} 
\left[ a_{*}(\vec{r}) \, E_{*}(t,\vec{x}) \, \Phi(t,\vec{x},\vec{r}) \right] 
\psi(\vec{x},\vec{r}) \: d\vec{x} \, d\vec{r}  
\end{equation}

By the Upwind Flux Rule, if $a_{*}(\vec{r}) \, E_{*}(t,\vec{x}) > 0$ in $\partial_{\varphi}^{\pm} \Omega_{I}$ 
, $q = 0, ...,5 \, ,$

{\small
\begin{eqnarray*}
(A_{6*}) 
&
\approx
&
\sum_{p=0}^{5} 
\left[ 
\int_{x_{i - \ot}}^{x_{i + \ot}} 
\int_{y_{j - \ot}}^{y_{j + \ot}} 
E_{*}(t,\vec{x}) \, \eta^{p}_{i,j}(\vec{x}) \, \eta^{q}_{i,j}(\vec{x}) \:dx \, dy 
\right] \times 
\\
& &
\left\lbrace 
\left[ 
\int_{r_{k - \ot}}^{r_{k + \ot}}
\int_{\mu_{m - \ot}}^{\mu_{m + \ot}}
a_{*}|_{n + \ot} \xi^{p}_{k,m,n}|_{n + \ot}  
\xi^{q}_{k,m,n}|_{n + \ot} dr d\mu 
%
\right. \right. 
\left. 
\left.
- \dfrac{2 \, \delta_{q3}}{\Delta \varphi_{n}} \int_{K_{kmn}}
a_{*}(\vec{r}) \xi^{p}_{k,m,n}(\vec{r}) d\vec{r} 
\right] \right. 
W_{I}^p (t)
\\
& & \left. 
- \left[ 
\int_{r_{k - \ot}}^{r_{k + \ot}} 
\int_{\mu_{m - \ot}}^{\mu_{m + \ot}} 
a_{*}|_{n - \ot} \, \xi^{p}_{k,m,n-1}|_{n - \ot} \, 
\xi^{q}_{k,m,n}|_{n - \ot} \: dr \, d\mu 
\right] 
 W_{ijkmn-1  }^p(t) \right\rbrace \, .
\end{eqnarray*}
}

If $a_{*}(\vec{r}) \, E_{*}(t,\vec{x}) < 0$ in $\partial_{\varphi}^{\pm} \Omega_{I}$ 
, $q = 0, ...,5 \, ,$

{\small
\begin{eqnarray*}
(A_{6*}) 
&
\approx
&
\sum_{p=0}^{5} 
\left[ 
\int_{x_{i - \ot}}^{x_{i + \ot}} 
\int_{y_{j - \ot}}^{y_{j + \ot}} 
E_{*}(t,\vec{x}) \, \eta^{p}_{i,j}(\vec{x}) \, \eta^{q}_{i,j}(\vec{x}) \:dx \, dy 
\right] 
\times 
\\
& & 
\left\lbrace 
\left[ 
\int_{r_{k - \ot}}^{r_{k + \ot}}
\int_{\mu_{m - \ot}}^{\mu_{m + \ot}} 
a_{*}|_{n + \ot} \, \xi^{p}_{k,m,n+1}|_{n + \ot} \, 
\xi^{q}_{k,m,n}|_{n + \ot} \: dr \, d\mu 
\right]
W_{ijkmn+1 }^p(t)
\right.
\\
& & 
 - \left[ 
\int_{r_{k - \ot}}^{r_{k + \ot}} 
\int_{\mu_{m - \ot}}^{\mu_{m + \ot}} 
a_{*}|_{n - \ot} \, \xi^{p}_{k,m,n}|_{n - \ot} \, 
\xi^{q}_{k,m,n}|_{n - \ot} \: dr \, d\mu 
\right. 
\left. \left. 
+
\dfrac{2 \, \delta_{q3}}{\Delta \varphi_{n}} \,  \int_{K_{kmn}}
a_{*}(\vec{r}) \,  \xi^{p}_{k,m,n}(\vec{r}) \: d\vec{r}  
\right] 
W_{I}^p (t)
\right\rbrace \, .
\end{eqnarray*}
}%

\end{document}